\documentclass[sigconf, nonacm]{acmart}
\usepackage{color}
\definecolor{Gray}{HTML}{ABD5FF}  
\usepackage{listings} 

\usepackage{mdframed}
\usepackage{array}
\usepackage{algorithm}
\usepackage{algpseudocode}
\usepackage{eqparbox}
\usepackage{caption}
\usepackage{subcaption}
\usepackage{tikz}
\usepackage{amsmath}
\usepackage{amsthm,thmtools}

\declaretheoremstyle[
  headfont=\bfseries,
  bodyfont=\itshape,
]{myplain}
\declaretheoremstyle[
  headfont=\bfseries,
  bodyfont=\normalfont,
]{mydefinition}
\declaretheoremstyle[
  headfont=\itshape,
  bodyfont=\normalfont,
]{myremark}

\definecolor{codegreen}{rgb}{0,0.6,0}
\definecolor{codegray}{rgb}{0.5,0.5,0.5} 
\definecolor{codepurple}{rgb}{0.58,0,0.82}
\definecolor{backcolour}{rgb}{0.95,0.95,0.92}

\lstdefinestyle{mystyle}{
    backgroundcolor=\color{backcolour},   
    commentstyle=\color{codegreen},
    keywordstyle=\color{magenta},
    numberstyle=\tiny\color{codegray},
    stringstyle=\color{codepurple},
    basicstyle=\ttfamily\footnotesize,
    breakatwhitespace=false,         
    breaklines=true,                 
    captionpos=b,                    
    keepspaces=true,                 
    numbers=left,                    
    numbersep=5pt,                  
    showspaces=false,                
    showstringspaces=false,
    showtabs=false,                  
    tabsize=2,
    lineskip=-1pt
}

\newcommand\vldbdoi{XX.XX/XXX.XX}

\newcommand\vldbvolume{14}
\newcommand\vldbissue{1}

\newcommand\vldbpagestyle{plain} 

\usepackage{xspace}    
\usepackage[labelfont=bf]{caption}    

\usepackage{caption}
\usepackage{subcaption}
\usepackage{tikz}
\usepackage{amsmath}

\usepackage{filecontents}

\usepackage{tikz}
\usepackage{amsmath}
\usepackage{subcaption}
\usepackage{pgfplots}
\usepackage{paralist}
\usepackage{multirow} 
\usepackage{enumitem}
\usepackage{float}
\usepackage{placeins}
\usepackage{graphicx}
\usepackage{booktabs}
\usepackage{multirow}
\usepackage{adjustbox}
\usepackage{caption}

\usepackage{makecell}
\usepackage{multirow}
\usepackage{graphicx}
\usepackage{colortbl}
\usepackage{tabularx} 

\usepackage[normalem]{ulem}
\useunder{\uline}{\ul}{}

\usepackage{xspace}    
\usepackage[labelfont=bf]{caption}    

\usepackage{filecontents}
\usepackage{amsfonts}

\mathchardef\h="2D

\usepackage{xurl}
\usepackage[htt]{hyphenat}
\usepackage{pythonhighlight}
\usepackage{hyperref}
\usepackage{cleveref}
\usepackage{listings}
\usepackage{outlines}
\usepackage{booktabs} 
\usepackage{array}

\newcommand{\remove}[1]{}

\newif\ifcomments
\ifcomments
    \providecommand{\ion}[1]{{\color{teal}{[ion: #1]}}}
    \providecommand{\joey}[1]{{\color{magenta}{[joey: #1]}}}

    \providecommand{\shu}[1]{{\color{red}{[shu: #1]}}}
    \providecommand{\simon}[1]{{\color{cyan}{[simon: #1]}}}
    \providecommand{\accheng}[1]{{\color{purple}{[accheng: #1]}}}
    \providecommand{\sarah}[1]{{\color{red}{[sarah: #1]}}}
    \providecommand{\danny}[1]{{\color{blue}{[danny: #1]}}}
    \providecommand{\gvernik}[1]{{\color{olive}{[gvernik: #1]}}}
    
    \providecommand{\michael}[1]{{\color{violet}{[michael: #1]}}}
    
    \providecommand{\souj}[1]{{\color{red}{[souj: #1]}}}

\else
    \providecommand{\ion}[1]{}
    \providecommand{\joey}[1]{}
    \providecommand{\shu}[1]{}
    \providecommand{\simon}[1]{}
    \providecommand{\gvernik}[1]{}
    \providecommand{\accheng}[1]{}
    \providecommand{\sarah}[1]{}
    \providecommand{\danny}[1]{}
    \providecommand{\michael}[1]{} 
    \providecommand{\souj}[1]{} 
\fi

\newcommand{\sys}{SkyStore\xspace}

\begin{document}

\date{}

\title{\sys: Cost-Optimized Object Storage Across\\Regions and Clouds}

\author{
  Shu Liu$^{1}$, Xiangxi Mo$^{1}$, Moshik Hershcovitch$^{2}$, Henric Zhang$^{1}$, Audrey Cheng$^{1}$}  
\author{
  Guy Girmonsky$^{2}$, Gil Vernik$^{2}$, Michael Factor$^{2}$, Tiemo Bang$^{1}$, Soujanya Ponnapalli$^{1}$} 
\author{
  Natacha Crooks$^{1}$, Joseph E. Gonzalez$^{1}$, Danny Harnik$^{2}$, Ion Stoica$^{1}$ 
}

\affiliation{
  $^{1}$UC Berkeley 
  $^{2}$IBM Research
}



\begin{abstract} 
Modern applications span multiple clouds to
    reduce costs, avoid vendor lock-in, and 
    leverage low-availability resources in another cloud.
However, standard object stores operate within a single cloud, forcing
    users to manually manage data placement across clouds, i.e.,
    navigate their diverse APIs and handle
    heterogeneous costs for network and storage.
This is often a complex choice:
    users must either pay to store objects in
    a remote cloud, or pay to transfer them over the network
    based on application access patterns and cloud provider cost offerings.
To address this, we present \sys, a unified object store that addresses
    cost-optimal data management across regions and clouds.
\sys introduces a virtual object and bucket API to hide
    the complexity of interacting with multiple clouds.
At its core,
    \sys has a novel TTL-based data placement policy that
    dynamically replicates and evicts objects according to application access patterns
    while optimizing for lower cost.
Our evaluation shows that
    across various workloads,
    \sys reduces the overall cost by up to 6$\times$ over academic baselines and 
    commercial alternatives like AWS multi-region buckets. 
\sys also has comparable latency, and its availability and fault tolerance
    are on par with standard cloud offerings. We release the data and code of \sys at  \href{https://github.com/skyplane-project/skystore}{{https://github.com/skyplane-project/skystore}}. 
\end{abstract}

\raggedbottom
\widowpenalty=100

\maketitle

\pagestyle{\vldbpagestyle}
\begingroup
\renewcommand\thefootnote{}\footnote{\noindent
This work is licensed under the Creative Commons BY-NC-ND 4.0 International License. Visit \url{} to view a copy of this license. For any use beyond those covered by this license, obtain permission by emailing \href{mailto:info@vldb.org}{info@vldb.org}. Copyright is held by the owner/author(s). Publication rights licensed to the VLDB Endowment. \\
\raggedright Proceedings of the VLDB Endowment, Vol. \vldbvolume, No. \vldbissue\ %
ISSN 2150-8097. \\
\href{https://doi.org/\vldbdoi}{doi:\vldbdoi} \\
}\addtocounter{footnote}{-1}\endgroup

\newcommand{\co}{\text{cost}_{\text{optimal}}}
\newcommand{\cp}{\text{cost}_{\text{P}}}
\newcommand{\teven}{T_{\text{even}}}
\newcommand{\tnext}{T_{\text{next}}}
\newcommand{\tev}{\text{TTL}}
\newcommand{\ttl}{\tev}
\newcommand{\uval}{U_{\text{perf-val}}}




\section{Introduction}
\shu{Incorporate real-world use cases: Add practical application examples or case studies, showcasing GlobalStore’s utility across specific industries (e.g., media, healthcare). Demonstrating real-world implementations will help underline its cost and data management benefits.} In the rapidly evolving landscape of cloud computing, applications increasingly span multiple regions and clouds. Organizations adopt multi-cloud software to reduce costs, avoid vendor lock-in, improve fault tolerance, increase the availability of specific capabilities beyond a single region or cloud, or support geo-distributed services~\cite{jain2022skyplane,cloudcast,skypilot}.
For instance, deploying a model serving service on multiple clouds reduces monetary costs by up to 50\% on low-availability resources (e.g., GPUs) compared to a single cloud~\cite{sky-serve}.
Today, these applications rely on object storage services (e.g., Amazon S3, Google Cloud Storage, Azure Blob Storage, and IBM Cloud Object Store \cite{aws-s3,gcp-storage,ms-azure-blob,ibm-cloud-storage})
to manage vast amounts of data.

Unfortunately, existing object stores operate within their respective clouds and typically limit their operations to specific regions; commercial systems like AWS and GCP only support multi-region but not multi-cloud replication \cite{aws-access-point}.
As a result, users manually handle data placement across clouds or regions, and their solutions cluster around two extremes: store locally or replicate everywhere. While storing all data in a single region simplifies data management and reduces storage costs, it increases egress expenses when data is accessed from another cloud region~\cite{cloudcast,aws-pricing,azure-pricing,gcp-pricing}. 
On average, data transfers across clouds cost 23$\times$ more compared to transfers within the same cloud.
On the other hand, replicating data to multiple regions and clouds~\cite{aws-replication, gcp-multi-region-bucket, minio, microsoftDataRedundancy}
may reduce network access fees but can significantly increase storage expenses. 
For instance, storing the training data for a Llama3 model with 15 trillion training tokens (60 TB in size)~\cite{llama}
    in AWS, GCP, and Azure standard storage buckets across different regions costs up to \$300K per month~\cite{aws-pricing, gcp-pricing, azure-pricing}.

A plethora of academic solutions have been proposed to address data storage in multi-region and multi-cloud settings.
However, these solutions are optimized to reduce latency~\cite{volley, nomad, pileus} and often ignore data transfer costs, which become prohibitive across multiple clouds. 
The most relevant work in this area is SPANStore~\cite{spanstore}, a multi-cloud storage system considering cost and latency tradeoffs. However, SPANStore does not account for replication costs and assumes that data access patterns do not change over time, significantly limiting its practical applicability. 

Consequently, there is a need for a multi-region, multi-cloud data placement solution that 
minimizes the total monetary cost for various cloud applications.
The key challenge in developing such a system is that cloud applications are highly diverse, and their data access patterns vary across several dimensions: object size, location distribution, and the recency and frequency of data accesses.
For instance, for applications that perform repeated reads, like model training, it is cheaper to 
    replicate data to accessed regions and avoid additional network costs for subsequent reads.
In contrast, for applications that read infrequently, like satellite imagery,
    it is more cost-effective to pay for network transfers occasionally. 

In this paper, we present \sys,
    a cost-optimized multi-cloud object store that adapts to the diverse and dynamic access patterns of applications.
\sys provides a single uniform API
    that emulates a local object store and 
    transparently manages data across
    clouds and regions while minimizing cost.
In a nutshell, \sys provides an \emph{overlay} cloud service on top of existing object stores that operate in specific regions and clouds. It decides where an object should be stored and the locations it should be replicated to, if at all, and consistently manages these data copies.

At a high level, \sys solves a caching problem: it decides to ``cache'' (i.e., replicate objects to their accessed cloud region) and to ``evict'' (i.e., remove object copies from cloud regions if they are unlikely to be re-accessed) based on application access patterns. Unlike traditional caches that optimize for performance and have a finite capacity, \sys minimizes monetary costs as capacity is virtually unlimited; it accounts for the non-trivial network and storage costs incurred to cache objects.

Accordingly, \sys needs to make two decisions: (i) when to cache (i.e., replicate) an object and (ii) when to evict an object. 
First,
\sys adopts a store-local and copy-on-read replication policy. 
When a user writes an object, \sys stores it in the local region to minimize cost and latency. If a user reads this object from another cloud region, 
\sys replicate it from an available region with the lowest transfer fees and ensures that network costs are only incurred for objects that are accessed. 

Second, \sys leverages a novel adaptive Time-To-Live (TTL)-based eviction policy that balances the cost of storing and transferring an object if the object is accessed again.
Our policy reduces costs by adapting the TTL assigned to objects based on the current workload and attempts to learn the optimal TTL for each replica. \sys captures past workload access patterns to estimate TTL costs and periodically updates these values while remaining robust to workload variance compared to traditional TTL-based methods (Section~\ref{s:eval}).
We also show in Section~\ref{sec:latency} how latency considerations can be incorporated into our cost-centric framework, generalizing it to hybrid clouds~\cite{ceph-RGW, minio} and clouds without explicit cost models~\cite{cloudflare-r2}.

We implement \sys to seamlessly integrate with existing clouds (e.g., Azure Blob, AWS S3, and GCS).
\sys offers a \textit{virtual bucket} and \textit{virtual object} abstraction via a standard S3-compatible API, allowing users to manage data as if all their data were in a single region.  
\sys provides the same consistency guarantees~\cite{aws-consistency, gcp-consistency, azure-consistency} as its underlying object stores and similar fault tolerance guarantees as existing services. 
We evaluate \sys on various workloads retrieved from IBM object store traces in the Storage Networking Industry Association (SNIA). 
Our prototype has comparable latency relative to the state-of-the-art systems, and our simulations show that \sys achieves up to 6$\times$ cost savings. In summary, our contributions are as follows: 
\vspace{-0.211em}
\begin{enumerate}
    \item We design a novel cost-optimized data replication policy that can adapt to diverse workload patterns in the multi-region, multi-cloud setting. 
    \item We implement \sys, a cost-efficient multi-cloud storage system that provides virtual object and bucket abstractions, seamlessly integrating with S3, GCS, and Azure Blob Storage as storage backends. 
    \item We evaluate against state-of-the-art policies, showing that \sys's policy can substantially reduce cost by up to 6$\times$ over SNIA object store traces \cite{snia-ibm-object-store-trace} compared to TTL-CC\cite{TTLbasedCC2019}, SPANStore\cite{spanstore}, and commercial systems like AWS Multi-Region Replication\cite{aws-access-point} and JuiceFS\cite{juicefs}. 
\end{enumerate}

\section{\sys Placement Policy} \label{s:policies}
We first provide an overview of \sys. \sys seeks to minimize dollar cost given a particular \textit{cloud pricing model} (Section~\ref{subsec:cloudcosts}) and \textit{modes of operations} (Section~\ref{subsec:modes}). To do so, it adopts an \textit{on-demand} approach to object placement and leverages a simple \textit{write-local} policy for data storage, which it combines with a \textit{read-driven} policy for data replication (Section~\ref{subsec:overview})

\shu{R1 / W1 (D1): if S3's default configuration is used alongside GlobalStore to replicate data across multiple regions, could this lead to significant data redundancy? For instance, in S3's default setup, an object is stored across multiple Availability Zones (AZs) within a single region. By replicating the object to other regions, S3's mechanisms will create additional replicas in those regions, potentially resulting in excessive duplication of data. While using S3's built-in cross-region replication, the number of replicas will be reduced.}
\subsection{Cloud Pricing Models}
\label{subsec:cloudcosts}
Cloud pricing consists of data storage, network, and operational charges \cite{gcp-pricing}. Most cloud vendors charge storage per GB per month based on the geographic region, provider, and storage class. For example, standard storage in \textit{gcp:southamerica-east1} costs $1.75\times$ more than S3 standard storage in \textit{aws:us-east-1}. The cloud provider also charges network (egress) costs based on the volume of data moved out of a particular cloud region \cite{aws-pricing}. This can differ by up to 15$\times$ within the same cloud and 19$\times$ between different clouds. Operations made to the cloud storage service are also charged: this cost is usually much cheaper than storage and network charges, with an average of 0.04 cents per thousand requests. Thus, we will mainly consider storage and network pricing in our discussion. 

\subsection{Modes of Operations}
\label{subsec:modes}
We explore two modes of object replication and eviction. In the \textbf{\textit{Fixed Base (FB)}} mode, each object has a designated primary region where its replica is never evicted. For example, data initially stored in AWS remains there permanently, while additional replicas are added or removed in other cloud regions based on demand. Alternatively, the \textbf{\textit{Free Placement (FP)}} mode allows replicas to be placed in any region, with the only requirement being that at least $k$ copies are always maintained (e.g., we explore $k=1$). 

\subsection{\sys Overview}
\label{subsec:overview}
\sys{}, as a multi-cloud storage system, must fundamentally answer these questions: where to \textit{write} objects, where to \textit{read} objects from, and how to \textit{replicate}. We briefly describe them in turn. \newline 
\textbf{Write Policy} \sys{} adopts a \textit{write-local} strategy. For data storage, \sys stores data in the region where the write request originates. This minimizes write latency and reduces egress costs for write operations, ensuring data is immediately available in the local region. In the fixed base mode, we set the base region of the object to the initial local write location. Consecutive write to the objects creates a new object with an updated version in its write location, where versioning is managed by \sys control plane (Section~\ref{s:arch-control-plane}). \newline 
\textbf{Read and Replication Policy}  \sys{} adopts a \textit{replicate-on-read} strategy.
Upon receiving a read request, \sys selects the cheapest region where the replica resides to retrieve the data and creates a local replica to optimize future reads. 
Replicate-on-read contrasts with proactive replicate-on-write methods used in AWS Multi-Region bucket and SPANStore \cite{aws-access-point, spanstore}, which pushes all data to a predicted set of regions upon write operations. Unless the prediction is accurate, such a model can lead to high egress costs and storage charges (Section~\ref{s:two-site-solution}). 
\sys reactively replicates to reduce future egress costs and performs eviction described in Section~\ref{s:evict} to keep storage costs in check. 

\section{\sys Eviction Policy}
\label{s:evict}
In this section, we discuss a cost-minimized cache eviction problem in a two-region base and cache setting (Section~\ref{s:problem-twosite}). We then introduce our cost-aware eviction policy (Section~\ref{s:two-site-solution}) and show how it extends to multiple regions and clouds (Section~\ref{s:multi}).

\subsection{2-Region, Base and Cache Problem}
\label{s:problem-twosite}

Consider a two-region setup: a base region storing all the objects that never get evicted, and a cache region that reads from base and replicates on read. 
We denote $S$ as the storage cost (\textbf{\$/GB*Month}) in the base region and $N$ as the egress cost (\textbf{\$/GB}) for moving an object over the network between the base and the cache region\footnote{For simplicity, we are ignoring the associated operation costs (e.g., cost for every PUT or GET) that are typically lower than the storage and egress costs.}. 
Aggressive replication in the cache region can lead to prohibitively high storage costs, especially as replicas accumulate over time. Thus, we now explore how to cost-effectively evict replicas in the cache region under this simple 2-region setup.

\subsubsection{The Clairvoyant Greedy Policy (CGP)}
\label{sec:clairvoyant}


We measure ourselves against a cost-optimal policy that is given access to an {\em oracle} that knows exactly when an object will be read in the future (if at all) in the cache region. This is akin to the Belady cache eviction algorithm \cite{belady}, but adapted to our problem setup. 
A key parameter in the clairvoyant strategy is the {\em break-even time}. This is the duration in which the cost of storing an object equals the cost of evicting it and fetching it again across the network (i.e., the storage cost equals the egress cost). We denote this as $\teven$, where:  
\begin{equation}
   T_{\text{even}} =  N  / S \label{eqn:teven}
\end{equation}

For example, we have $\text{Cost}_\text{storage} = \$0.026 \text{ per GB per month}$ for \texttt{aws:us-west1} and $\text{Cost}_\text{egress} = \$0.02 \text{ per GB}$ between \texttt{aws:us-east1} and \texttt{aws:us-west1}. Thus,  
    $T_{\text{even}} \approx 0.77 \text{ months}$ for the edge between these two regions.\footnote{Prices taken in Sept. 2023.}

Since the eviction of one object is independent of others, it is clear that the best one can do is to cache an object as long as the cost of storage is lower than the cost of bringing it again over the network and vice versa. 
Every time an object is read, the clairvoyant policy accesses an oracle that returns the time duration until this object will next be read. We denote $T_{\text{next}}(o,i)$ as the time between the $i^{th}$ and ${i+1}$ reads of object $o$. The strategy then compares $T_{\text{next}}$ with the break-even time $T_{\text{even}}$ and decides whether to evict the object. An object with no next GET is immediately evicted. In summary, the clairvoyant policy upon the $i^{th}$ access to object $o$ works as follows:
\vspace{-0.2em}
\[Clairvoyant(o,i) = \left\{
  \begin{array}{lr}
    \text{evict :} &  T_{\text{next}}(o,i) > T_{\text{even}}\\
    \text{keep :} &  T_{\text{next}}(o,i) \leq T_{\text{even}}
  \end{array}
\right.
\]

\remove{
\begin{equation}
\text{Cost} = \text{storage cost} = ( T_{\text{next}}(o,r) ) \times \text{Size}(o) \times \text{Cost}\_\text{storage}(r)
\end{equation}

If an item is evicted, then its egress cost is paid for the next access,

\begin{equation}
\text{Cost} = \text{next access cost} = \text{Size}(o) \times \left( \frac{1}{m} \right) \sum_{j=1}^{m} \text{Cost}_\text{egress}(r_j, r)
\end{equation}

The break-even point is when the cost of evicting an object would be the same as storing it. It is defined by:  
\begin{equation}
   T_{\text{even}} =  \text{Cost}_\text{egress} / \text{Cost}_\text{storage}(r)
\end{equation}
}
\subsubsection{The $\teven$-policy}
\label{s:teven-polcy}
A simple policy ($\teven$-policy) will be setting $\tev$ to the break-even time $\teven = \frac{N}{S}$ and refresh upon each access. It has the following properties:

\begin{enumerate}
    \item The cost of the $\teven$-policy is at most twice the clairvoyant policy.
    \item $\forall$ eviction policy $\exists$ a workload for which the policy costs twice as much as the clairvoyant policy. 
\end{enumerate}

\paragraph{Proof for (1)} The cost per GB for a single object under the optimal clairvoyant policy includes paying network cost $N$ for initial GET, storage cost $\tnext(i) \cdot S$ for storing the object until the next access, and network cost $N$ for re-fetching the object after eviction:
$$ N + \sum_{i| \tnext(i) \leq \teven} \tnext(i) \cdot S + \sum_{i| \tnext(i) > \teven} N $$

The cost of the $\teven$-policy policy is:
$$ N  + \sum_{i| \tnext(i) \leq \teven} \tnext(i) \cdot S + \sum_{i| \tnext(i) > \teven} (\frac{N}{S} \cdot S + N) + \frac{N}{S} \cdot S$$
The first two parts are identical. However, for $\tnext(i) > \teven$, $\teven$-policy needs to pay additional storage cost until the break-even point$\frac{N}{S} \cdot S$, evict it, and pay for network cost to re-fetch. The last $\frac{N}{S} \cdot S$ accounts for keeping this object around after its last GET. 
Thus, $\teven$-policy is bounded by $2\times$ that of the optimal. 

\paragraph{Proof for (2)} 
We claim that for any eviction strategy, an adversarial workload exists that costs more than twice that of the optimal strategy.
Consider a single object. After its first access, the eviction policy $P$ must decide when to evict. If $P$ decides to evict the object after more than $\teven$ time, then the workload never asks for this object again. In such a case, the optimal cost ($\co$) is just the initial GET cost, $N$, while the cost for policy $P$ ($\cp$) is greater than $N + \teven \cdot S = 2N$, double the optimal. 

Alternatively, if policy $P$ evicts the object earlier (at $t < \teven$), the workload issues a new GET shortly after $t + \varepsilon$, where $t + \varepsilon<\teven$. 
The optimal cost $\co = N+(t+\varepsilon)\cdot S$ in this case , while the $\cp$ becomes $2N+t\cdot S$.
Since the object is in the cache again, this process can repeat.
After $k$ iterations, we have $\co = N+(\sum (t_i+\varepsilon))S$ whereas $\cp = (k+1)N + (\sum t)S$.
Their difference grows as $\cp - \co = kN + k\varepsilon$. For large enough $k$, since $\co <(k+1)N$, $k \varepsilon> N$ will cause $\cp-\co > \co$ and give us a ratio $\geq 2$.  


These two properties demonstrate that if nothing is known apriori about the workload, then the $\teven$-policy is the safest strategy one could hope for.
However, distributions are not chosen adversarially in reality. Quite a bit can be learned about the distributions of the workload at hand, which should be used to reduce costs. 

\subsection{\sys 2-Region Base \& Cache Eviction}
\label{s:two-site-solution}
Now we discuss how \sys policy learns workload distributions over time and aims to set the cost-optimal $\tev$ for objects in the 2-region setup.
\subsubsection{TTL-based eviction.}

Inspired by $\teven$, \sys assigns each replica (i.e., a copy of an object) a TTL (Time To Live) value. For the free-placement (FP) mode, if the replica is not accessed within TTL time, it will be evicted as long as it is not the sole remaining replica. 
In fixed-base (FB) mode like the 2-region setup (Section~\ref{s:two-site-solution}), the replica can only be evicted if it is not in the base location.  

There are two common TTL-based eviction methods. The first, used in CDNs~\cite{ttl-cdn}, invalidates a cached object after its TTL expires, regardless of access frequency, to prevent stale data.
\sys takes a different approach and resets the TTL on each access to reduce network costs and avoid evicting frequently accessed objects. \sys eviction policy periodically scans and evicts objects that have not been accessed within $\tev$ time. The policy for evicting object $o$ at region $R$ is as follows: 
\[ Policy(o, R) = \left\{
  \begin{array}{ll}
    \text{evict :} &  \text{time since last access} > \tev(o, R) \\
    & \ \& \text{ not sole copy}\\
    \text{keep :} &  \text{otherwise}
  \end{array}
\right.
\]

\subsubsection{Adaptive TTL}\label{s:tevict}

The crux of \sys's approach is setting the right TTLs for replicas.
The main statistic we use to adapt $\tev$ is the time between accesses of objects, represented as a distribution of $\tnext$ at the cache region. 
We build a weighted histogram where each cell corresponds to a time range, and the weight reflects the total size of GETs within $\tnext$ in that range. This histogram is collected per region per workload. We define relevant notations in Table~\ref{tab:notations}. The value in each histogram cell is denoted as:
$$hist(j)  = \sum_{o, i| \tnext(o,i) \in range(j)} size(o)$$
This histogram accounts for all re-reads in the cache region for the workload. However, it does not account for what happens to objects after their last access. For this, we use an additional histogram called $last$ to track the latest access time. Given these histograms and a TTL value, we compute the expected cost for the $\tev$ as: 
\vspace{-0.4em}
\setlength{\jot}{0.5pt} 
\begin{align*}
\scriptsize
ExpectedCost(\text{$TTL$}) = 
&\sum_{\substack{o \in R \\ \text{Requested}}} Size(o) \cdot \mathbb{I}[\text{Fetched remotely}] \cdot N\\
&+ \sum_{\substack{j \in hist \\ t(j) \leq \tev}} hist(j)\cdot \widehat{t}(j) \cdot S \\
&+ \sum_{\substack{j \in hist \\ t(j) > \tev}} hist(j) \cdot (N + \tev \cdot S) \\
&+ \sum_{\substack{j \in last}} last(j) \cdot \tev \cdot S
\end{align*}

\begin{table}[t]
    \centering
    \begin{tabular}{ m{6.5em}  m{15em}}
      \toprule
      {\bf Parameter} & {\bf Description}  \\
      \midrule
      $range(j)$ & Time interval of the $j^{th}$ cell \\
      $t(j)$ & Maximum time in $range(j)$ \\
      $\widehat{t}(j)$ & Mean time in $range(j)$ \\
      $hist(j)$ & Bytes re-read after time $t \in range(j)$ \\
      $last(j)$ & Bytes not read in $t \in range(j)$ \\
      \bottomrule
    \end{tabular}
    \vspace{0.5em}
    \caption{Eviction parameters in \sys.}
    \label{tab:notations}
\end{table}

The first term accounts for the initial read cost of all objects requested from the region $R$, using an identity function over remote reads. If it is a local read, the cost would be 0; if fetched remotely, the cost would be $N$. The second accounts for hits -- objects that are re-read and exist in the region. The third accounts for misses -- objects that are evicted and brought into the region with additional network cost, and the last term accounts for storage costs of objects that have not yet been re-read.
We iterate over possible $\tev$ values at the same granularity as the histogram and select the one with the lowest expected cost. 

\begin{figure}
    \centering
    \includegraphics[width=.85\linewidth]{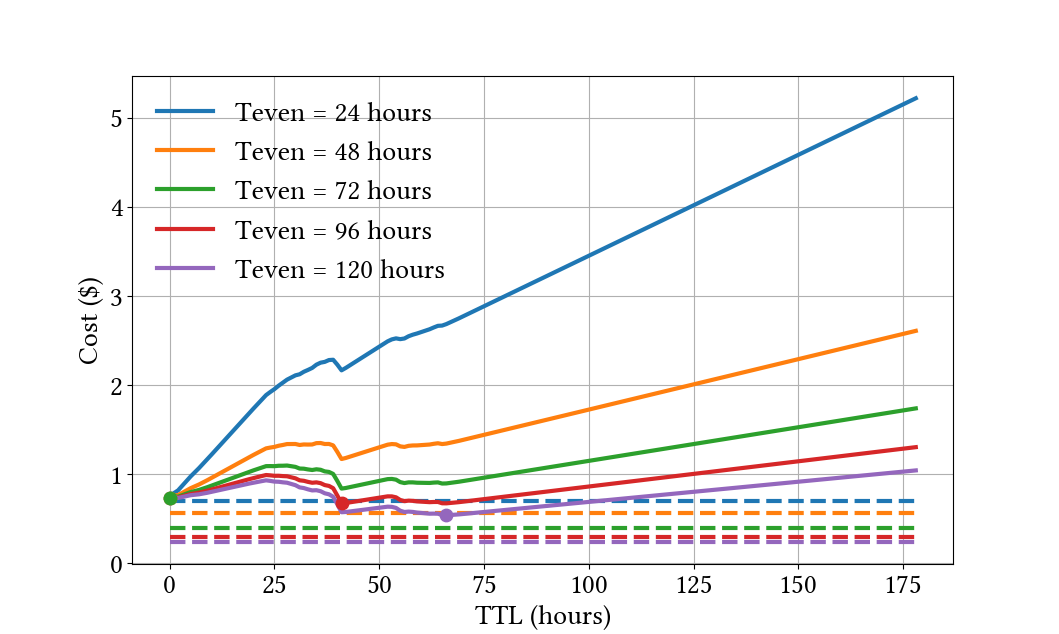}

    \vspace{-1em} 
    \caption{The expected cost as a function of TTL on a trace with an hourly histogram. The dashed line shows the optimal policy cost, and the dot marks the minimum cost point.}
    \label{fig:costhist}
    \vspace{-1em}
\end{figure}

The best TTL chosen is influenced by the specific workload and the network and storage costs.  Figure~\ref{fig:costhist} shows an example of the expected cost as a function of TTL for an IBM trace with different pricing choices. A lower value of $\teven$ indicates that storage costs are higher (relative to the network costs) and means that shorter TTLs would fare better, as seen in the example.   

\shu{R1 / D2: Section 3.2.2 discusses Adaptive TTL. However, the paper lacks sufficient detailed description of the histogram. What is the purpose of the histogram? What does the composition of the ranges (j) look like? What is the extent of the ranges, and how is it maintained? etc. Using hist(j) to estimate an appropriate TTL? How accurate can the calculations based on hist(j) be?}

\subsubsection{Granularity of Histogram} 
In prior discussions, we collected a histogram to study cache region access distribution. In object stores, bucket-level granularity reliably reflects workload access patterns, as bucket distributions remain stable over time. Object-level statistics, however, can be misleading. For instance, in one IBM trace, there are bursts of 2-8 consecutive GETs to the same object within 10 minutes of each other, followed by no further access to that object. Methods that focus on learning each object's pattern separately~\cite{ewma} or assume Poisson-like distributions~\cite{TTLbasedCC2019} fail to capture this bursty behavior. \sys generalize bucket-level patterns and assign a TTL that ensures replicas remain available during bursts but are evicted soon after.
\shu{R1 / D3: In 3.2.3, "bucket-level granularity reliably reflects workload access patterns, as bucket distributions remain stable over time" Why the distribution of barrels remains stable over time?}

The granularity of the histogram is also directly related to our possible choices for setting the TTL. A more granular histogram gives us additional information and allows us to choose TTLs more accurately, achieving better cost savings. On the other hand, a large histogram burdens the memory and computation requirements of the system. Recall that the histogram should potentially cover a time duration of many months, yet at times, the best eviction policy calls for evicting objects within minutes or even seconds. 
To balance this tradeoff, we support a variable range for histogram cells and attempt to have high granularity for small TTL values and low granularity for larger ones. For the first minute, we use a per-second granularity (taking up 60 cells). Beyond that, we employ a logarithmic base granularity with a low base of $1.02$. This ensures that the ratio between two consecutive potential TTL values is no more than 2\%. In turn, the difference in storage cost between two consecutive TTLs is also bounded at 2\% as the cost is linear in the time the replicas are stored. 
Using 740 cells at this log granularity covers $(1.02)^{740}$ minutes, which amounts to almost 2 years. An additional 60 cells cover the first minute, and we thus manage to cover nearly 2 years with an 800-cell histogram. 

To account for changing workload distributions and application behavior over time, we opt to periodically collect a new histogram (while still keeping the previous histogram). Once the new histogram has a sufficiently long history, the old histogram can be discarded. Our investigations indicate that the histogram should be longer than the $\teven$ time to be effective. 
\shu{R1 / D4: What is the granularity of replication/eviction? Is it at the individual object level or the entire bucket level? When the granularity of replication is at the object level, is the statistical accuracy of the bucket-level data maintained? Conversely, when replication is at the bucket level, does it lead to the replication of many unnecessary objects, resulting in network waste?
}
\subsection{\sys Multi-Region Eviction}\label{s:multi}

\subsubsection{Choosing adaptive TTLs.} 

\begin{figure}
    \centering
    \includegraphics[width=.85\linewidth]{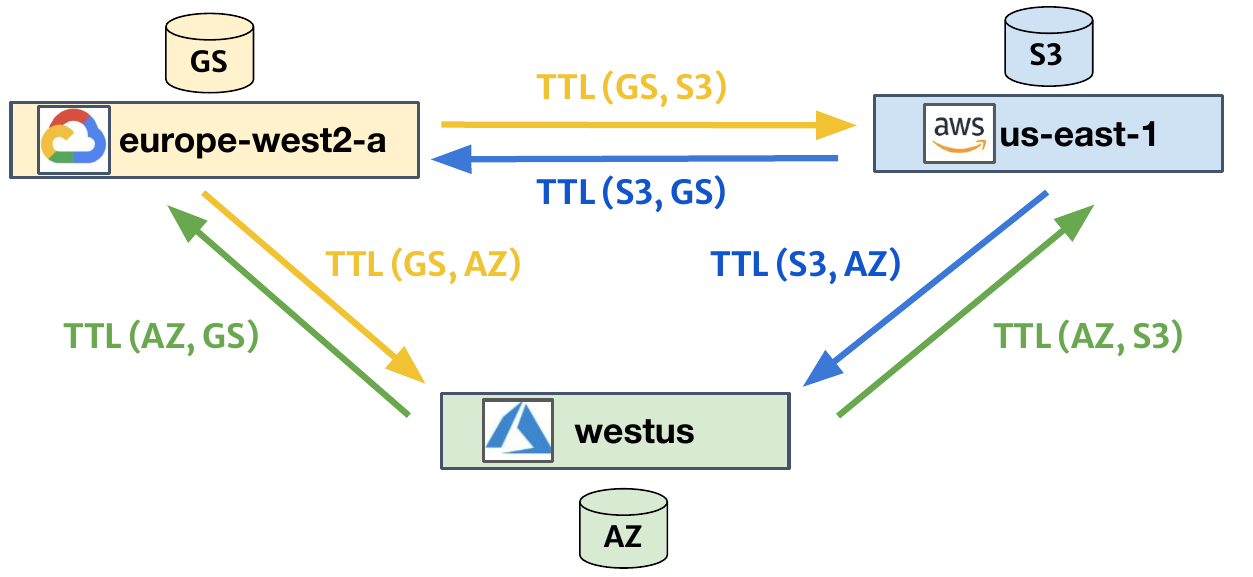}
    \vspace{-1em} 
    \caption{The multi-cloud setting as a directed graph with a TTL assigned per each directed edge.}
    \label{fig:ttl}
\end{figure}

We tackle the multi-region setting by breaking the problem into a pairwise problem similar to the 2-region setup; then, we set the TTL for each pair of source and target regions. Namely, we view the multi-region setting as a fully directed graph where each node is a region, and for each directed edge, we compute a TTL corresponding to this edge (as shown in Figure~\ref{fig:ttl}).

The TTL assigned to an object at a specific region is then deduced from the TTLs of the edges directed at this region. 
We denote $\tev(R_i,R_j)$ as the chosen TTL value for the edge from region $R_i$ to region $R_j$ and $\tev(o, R_j)$ as the TTL assigned to an object $o$ at region $R_j$. The eviction TTL of an object depends on the relevant regions that hold a replica of the object $o$. The TTL of the object at each region is then chosen to be the minimal TTL of edges from all such relevant regions.  Namely: 
$$\tev(o, R_j)= \min_{i | o \in R_i} \tev(R_i,R_j) $$

The TTL of an edge is assigned as a function of the incoming network cost, so the cheaper the cost, the lower the TTL. Since we use the cheapest available source in case of a cache miss, this corresponds to the minimal TTL.  Our method for calculating an edge's TTL is detailed in Section~\ref{s:two-site-solution}: we take the storage cost at the target, the network cost from some source region to the target, and statistics histograms of the workload in the target region as input. 
This final component is what makes our choice of TTLs adaptive. As time goes by, we learn from the access patterns of the workloads and change the associated TTLs accordingly.


Our approach assigns a local TTL to each object, which is the minimum of all relevant edge TTLs where the source region has replicas. This assumes a remote replica will still exist after the local TTL expires, enabling cost-efficient retrieval. However, since TTLs are set independently, this assumption may not always hold. To ensure correctness, we filter out cases where the local TTL plus storage start time exceeds the remote replica's eviction time, calculated as the replica's start time plus its TTL. This prevents reliance on replicas that may already be evicted.

\subsubsection{Latency Considerations}
\label{sec:latency}
A cost-centric policy could implicitly model performance, as resources often have associated price tags.
However, we observe that incorporating latency into a cost-driven framework is particularly challenging since it requires assigning a cost value to read performance, which is specific to users, applications, and objects. 

We propose a potential solution to model the price of cache hits and ask how much a customer is willing to pay for cache hits. 
Namely, if all objects are equally important, 
how much cost would the user be willing to pay for additional low-latency local read? 
We denote this value as {\em user performance value} or $\uval$, in dollar cost per byte.
We incorporate this into our methodology of carefully choosing a $\ttl$ as follows:
After finding the value of $\ttl$ that promises the lowest expected cost, we check if there is a higher $\ttl$ value for which the $\uval$ bounds the average cost per additional cache hit. 
More formally, if $\ttl'$ represents the eviction time that achieves the lowest expected cost, we choose the highest $\ttl$ value such that 
$$ \frac{ExpectedCost(\ttl) - ExpectedCost(\tev')}{\text{object byte count between $\tev$ and $\tev'$}} \leq \uval$$

In this model, users pay for objects until they are evicted, and their TTLs are reset upon the next access. We plan to compare the effectiveness of this approach to \sys's cost-centric policy and estimate its cost and latency tradeoffs in the future.


\section{\sys Architecture} 
\label{s:impl}
\label{sec:Architecture}
Building a cost-efficient multi-cloud object store requires addressing several key challenges. 
Such an object store must offer 1) a cohesive view of global objects stored across multiple regions and clouds 2) consistency across clouds and regions. Currently, consistency is typically guaranteed only within single-region object stores. 3) reliable data recovery in the event of failures, with guarantees comparable to single-region object stores. 

\begin{figure}
    \centering
    \includegraphics[width=.85\linewidth]{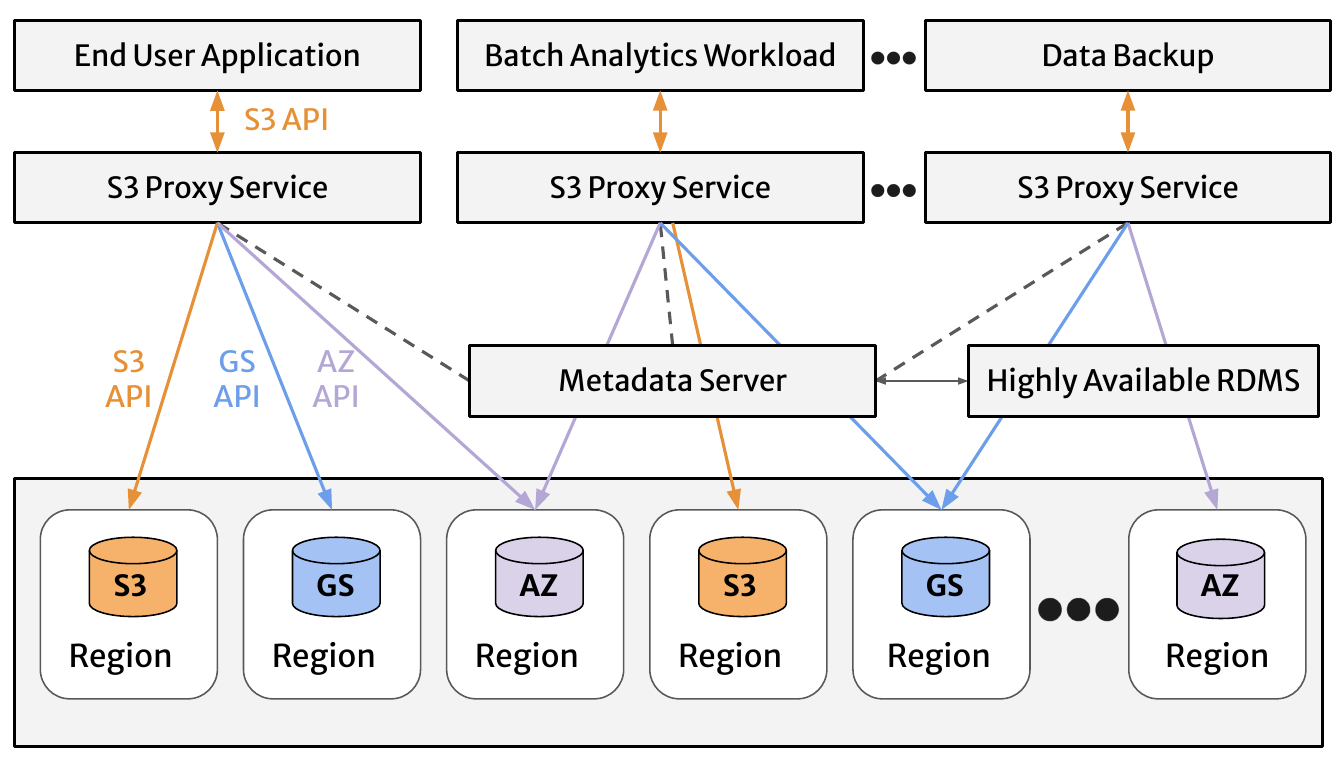}
    \vspace{-1em} 
    \caption{The system architecture of \sys. \shu{R1 / D7: Missing legends: need a sequence of steps of each request}
    }
    \label{fig:architecture}
\end{figure}

\sys is designed as an overlay layer on top of existing cloud object storage systems, including AWS S3~\cite{aws-s3}, Google Cloud Storage~\cite{gcp-storage}, and Azure Blob Storage~\cite{ms-azure-blob}. 
It consists of a client proxy service and a control plane, as shown in Figure~\ref{fig:architecture}. 
The client proxy fetches objects and supports the AWS S3 wire protocol~\cite{s3-api}, allowing users to seamlessly port applications using the S3 interface. The control plane, a stateless web server backed by a database, tracks object locations and redirects requests across cloud regions. We elaborate on the design of these components in Sections~\ref{s:api}-\ref{s:dataplane}. We then summarize \sys's consistency guarantees~(\S\ref{s:consistency}), and fault tolerance~(\S\ref{s:fault}).

\subsection{API: Virtual Object \& Bucket Abstraction} \label{s:api}
In object stores, objects are binary blobs identified by a key within a specific bucket. A bucket, serving as a namespace, is a collection of objects and the unit for placement and permission management. Traditionally, objects and buckets are confined to specific regions and clouds. As such, clients need to know the location and cloud of an object before accessing it. 
\sys abstracts this away with \textit{virtual object} and \textit{virtual bucket} that appear global to the user, with their physical locations managed transparently by \sys. This abstraction simplifies interaction with diverse cloud APIs by leveraging common concepts across providers. 
Users manage and access objects as if they were local, while \sys efficiently handles the routing and storage of these objects across different regions and clouds.
\subsection{Control Plane: \sys Metadata Server}
\label{s:arch-control-plane}
The \sys metadata server acts as the central coordinator for routing requests across multiple regions and clouds. Importantly, the control plane does not handle actual object data, eschewing any potential bottlenecks.
The metadata stored for each virtual object includes key information such as object size, last modified time, entity tag, and version ID. \sys also manages the mapping between virtual objects and their physical locations in each cloud region. A key component of this system is the \texttt{Policy} interface, which determines where to store objects on PUTs and where to fetch them on GETs. This interface supports various placement and eviction policies described in Section~\ref{eval:baselines}. 

\textit{Eviction Process} The metadata server collects $T_{next}$ statistics into histograms to assist with \sys decision-making. 
A background process runs periodically (once per day) to scan for objects exceeding their TTLs and initiates \texttt{DELETE} requests in the respective cloud object stores. 
This process is computationally lightweight since it only involves handling metadata, with the actual deletion handled by the cloud providers, so no data transfer occurs. 
In practice, this method incurs minimal overhead, as shown in Section~\ref{eval:end-to-end}. 
Alternatively, configuring lifecycle policies~\cite{aws-lifecycle} for objects in each bucket could remove the need for \sys to track TTLs, although these policies are typically limited to 1000 rules per bucket.

\shu{R1 / D6: The execution flow of the system is not sufficiently clear. When does the S3 Proxy Server need to interact with the Metadata Server? If processing each request requires communication with the Metadata Server, could the Metadata Server become a performance bottleneck? Furthermore, will the need for additional cross-region communication each time reduce the performance of user data reads (additional latency)? If not, how does the S3 Proxy Server know whether the replicated data has been deleted in the current region, and what measures are taken after the data is deleted?}



\subsection{Data Plane: S3-Proxy}
\label{s:dataplane}
The data plane handles user requests by interfacing with physical object stores through a \textit{S3-Proxy}. Requests are processed according to AWS S3 protocols~\cite{s3-api}, which we choose to implement due to its widespread popularity and market share~\cite{aws-market-share}. 
We support 14 common object store operations, including \texttt{create}, \texttt{delete}, \texttt{list} of buckets, and \texttt{head}, \texttt{get}, \texttt{put}, \texttt{delete(s)}, \texttt{list}, \texttt{copy}, and \texttt{multipart-upload} related operations. In our experience, this is sufficient to support almost all cloud workloads.
Upon accepting a request, the proxy reaches out to the \sys metadata server to learn the physical object store to issue this request to and then communicate with various cloud storage providers to fulfill it. 
The stateless design of our S3-Proxy ensures horizontal scalability. 
\begin{figure*}[t] 
    \centering
    \includegraphics[width=\textwidth]{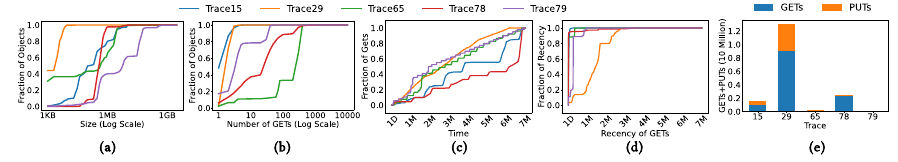}
    \vspace{-2.3em}
    \caption{\textbf{Trace Analysis}: We showcase object sizes (a), access frequency (b), burstiness or the fraction of GETs over time (c), the recency of GETs (d), and the PUT to GET ratio (d), for five representative IBM traces. We represent time in days (D) and months (M).}
    \label{fig:5_trace_new_analysis}
    \vspace{-2em}
\end{figure*}

\subsection{Consistency} 
\label{s:consistency}

\sys matches the consistency guarantees provided by existing object stores. Specifically, it 
supports both Read-After-Write Consistency (offered by S3) and Eventual Consistency \cite{aws-consistency, gcp-consistency, azure-consistency}. \newline 
\textbf{Read-After-Write Consistency} ensures that after a write (PUT), the latest version of an object is immediately available for reads (GET). This model is critical for applications requiring fresh data, such as e-commerce systems~\cite{strong-raw-consistency}. Our write local policy in Section~\ref{s:policies} provides the same read-after-write guarantees as a single region object store.
For cross-region access, with versioning enabled, \sys tracks both virtual object versions and the corresponding physical copies, directing reads to regions holding the most recent version.
Without versioning, \sys uses a Last-Writer-Wins policy, where the most recent write globally overwrites earlier copies in different regions, and synchronous replication ensures all regions are updated before the write is finalized.
\newline 
\textbf{Eventual Consistency}
allows faster, local reads by serving possibly stale data while updates propagate asynchronously across regions. This model minimizes network overhead and boosts access speed when real-time consistency is not required, making it ideal for use cases like backup systems and non-critical analytics~\cite{eventual-consistency}. In \sys, under Eventual Consistency, reads can be fulfilled from local or nearby regions even if they do not yet contain the latest version. Over time, updates are propagated to ensure that all regions eventually receive the most current version of the object.



\subsection{Fault Tolerance}
\label{s:fault}
\shu{Discuss GlobalStore’s fault tolerance and its cost:
- Describe recovery mechanisms to achieve reliability for high-stakes applications.
- What is the cost of maintaining fault tolerance, in particular for maintaining replica consistency in scenarios with frequent updates?
}
To ensure high availability and fault tolerance, \sys store-server can be run on a highly available database system such as Postgres with Primary-Secondary replication \cite{postgres-replication}. The metadata server regularly backs up data to cloud storage, allowing recovery during server failures. 
A secondary server in another region can use this backup if the primary server fails. 
Object data is durably stored in cloud object stores, so even during metadata failures, users can still locate and access objects across regions. In case of incomplete checkpoints, users can manually scan object stores to reconstruct missing metadata, ensuring no data is permanently lost.

To handle potential data plane failures, such as S3-Proxy or network disruptions, \sys implements a two-phase commit protocol \cite{two-phase-commit} to prevent metadata and object data corruption.
When a client issues a write request, the metadata server logs the intended action, committing it only after the object is successfully written to the cloud storage. If an error occurs, the metadata server rolls back the changes to maintain consistency, and uncommitted mutations are timed out. While this protocol ensures data integrity, it may introduce some performance overhead (Section~\ref{s:system-overhead}), particularly in high-throughput scenarios, due to the need for synchronous coordination between metadata and cloud storage. 
\shu{R1 / W2: The paper primarily focuses on read optimization by enhancing user access through the addition of local replicas. However, it ignores the costs associated with maintaining consistency among multiple replicas in scenarios with frequent updates.} 

\shu{R2: An analysis of GlobalStore’s fault tolerance under extreme failure scenarios (e.g., provider outages or network failures) would offer valuable insights. Describing its recovery mechanisms and illustrating how they work in real-time could improve readers' understanding of its reliability and potential for high-stakes applications. Is being fault tolerant implies an economic cost?}

\section{Implementation}
We prototype \sys as described in Section~\ref{sec:Architecture} to compare the end-to-end latency of \sys against other policies. 
\sys's metadata server is implemented
 in 3.5K lines of Python code to support various policies. It stores the metadata in a Postgres database by default~\cite{postgres-replication} and can be configured to support an SQLite backend~\cite{sqlite}.
The S3-proxy is implemented in about 9k lines of Rust code, connecting to AWS S3~\cite{aws-s3}, Google Cloud Storage~\cite{gcp-multi-region-bucket}, and Azure Blob Storage~\cite{ms-azure-blob}. In our experiments, we host the metadata server on \texttt{m5d.8xlarge} instance in \texttt{aws:us-east-1}. It contains 32 vCPUs, 128 GiB of memory, and 2 x 600 GB NVMe SSDs.
We instantiate a S3-proxy on each client VM that uses \texttt{m5.8xlarge}, \texttt{Standard\_D32ps\_v5}, and \texttt{n2\-standard\-32} instance types on AWS, Azure, and GCP, respectively. These client VMs contain 32 vCPUs, 128 GiB of memory, and 64 GB of storage. 

We also implement the \sys policy and all baselines outlined in Section~\ref{eval:baselines} in 1.9k lines of Python code to estimate the total cost of each of these policies across traces. Our simulations are run on a standard VM like \texttt{n4-standard-4} with 4 vCPUs, 16GB memory, and 32GB of storage.

\section{Evaluation} \label{s:eval}
In our evaluation, we answer the following questions: 
\begin{enumerate}[noitemsep=*,leftmargin=*]
    \item What are the cost benefits of \sys's replication policy across two regions within a single cloud? 
    \item Do the cost savings from \sys's policy scale to multiple regions across multiple clouds? 
    \item What are the end-to-end latency and cost savings of \sys in a real multi-cloud deployment? 
\end{enumerate}

We describe
    multi-cloud workloads (Section~\ref{eval:workloads-baselines-settings}), experiment setups, and our comparison baselines (Section~\ref{eval:metrics_baselines_deploy}).
We then showcase the cost improvements of 
    \sys against other baselines across two regions 
    within a single cloud (Section~\ref{eval:two-region}).
Then, we analyze the multi-cloud performance of \sys
    with three regions across three clouds (Section~\ref{eval:three-region}),
    and discuss its scalability to nine regions across the same clouds (Section~\ref{eval:nine-region}).
Finally, we discuss the end-to-end latency and cost of our \sys prototype
    on a real deployment across three regions and three clouds (Section~\ref{eval:end-to-end}) and conclude by measuring its overheads (Section~\ref{eval:overheads}). 

\subsection{Workloads: Multi-Region and Multi-Cloud}
\label{eval:workloads-baselines-settings}

We describe the object store traces we use
    (Section~\ref{eval:workload-gen}), outline their diverse
    characteristics (Section~\ref{eval:trace}), and 
    discuss our methodology to carefully generate 
    multi-cloud workloads from these traces.
This step is necessary as there are no publicly available multi-cloud traces to the best of our knowledge.

\subsubsection{Workload generation from traces}
\label{eval:workload-gen}
Our workloads are drawn from the SNIA
    IBM Object Store traces~\cite{SNIA_TRACE}.
These traces record a week of RESTful operations (e.g., GET, PUT, HEAD, DELETE) 
    for a single region within the IBM cloud~\cite{LRUvsFIFO}.
These traces effectively capture diversity across
    various dimensions: object sizes, recency, and frequency of accesses (as detailed in Section~\ref{eval:trace}).
However, object stores are typically designed for long-term data retention where objects are stored for several months to years~\cite{long-term-storage}. 
Since these short, week-long traces inadequately capture the life of objects in the cloud, we expand a day in each trace to a month for single cloud experiments and to three months for multi-cloud settings without changing their inherent characteristics like read-to-write ratio or request distributions. 

We pick five representative traces with salient characteristics
    in recency, frequency, size, burstiness, and PUT and GET distributions.
We outline their characteristics and the key insights that inform the generation of multi-region and multi-cloud workloads. 
In the interest of space, we use multi-region, multi-cloud workloads generated from these representative traces for all of our experiments.

\subsubsection{Trace characteristics}
\label{eval:trace}
Cloud applications have unique access patterns
    across dimensions like object sizes,
    PUT to GET ratios, access frequency, recency, and burstiness,
    as shown in Figure~\ref{fig:5_trace_new_analysis}.
We summarize these characteristics in Table~\ref{tbl:trace_characteristics}.

\begin{itemize}[noitemsep=*,leftmargin=*]
\item
\textit{Object sizes}: We categorize objects in four size ranges: tiny ($<$1KB), small (1KB to 1MB), medium (1MB to 1GB), and large (>1GB).
As seen in Figure-\ref{fig:5_trace_new_analysis}a,
    most of the objects accessed are small or medium in size,
    some are tiny, and very few are large.
Most traces have $<$0.5\% of tiny objects except two traces (T29 and T65)       with 30--45\% of tiny objects.
All traces have $>$35\% of small objects and notably, three traces
    (T15, T29 and T78) have a majority (56--97\%) of small objects.
About 34\% and $>$60\% of objects in T65 and T79 are medium-sized, while less than 20\% of the other traces have medium-sized objects.
None of the traces have large objects except T65 and T79, which rarely have large objects (<0.4\%).

\item
\textit{Access frequency and one-hit wonders}: 
    We categorize objects accessed in our traces as
    one-hits (1 GET), cold (1-10 GETs),
        warm (10-100 GETs), hot (100-1000 GETs), and super hot (>1K GETs).
As seen in Figure~\ref{fig:5_trace_new_analysis}b,
    our traces significantly differ in their distribution of repeated reads.
Two traces (T15 and T29) are almost entirely composed of objects that are 
    one-hits (98\% and 2\% respectively) or cold (52\% and 98\% respectively).
In contrast, T78 has a majority ($>$51\%) of warm objects and
    T65 has a majority ($>$67\%) of hot objects.
None except two traces (T65 and T78) have
    super hot objects in very small proportions ($<$0.1\%).

\item
\textit{Burstiness}: 
We define burstiness as the fraction of GETs over time.
As seen in Figure~\ref{fig:5_trace_new_analysis}c,
    our traces have distinct burst patterns over time with different spikes.
While one trace (T15) has an even distribution of accesses and has no accesses in the last two months,
    another trace (T78) has a burst with 60\% of GETs within the last two months.
In the rest of the traces, about 50\% of accesses arrive in the last two months.
Three traces (T29, T65, and T79) nearly have an equal distribution of GETs, with a noticeable spike, where 30\% of objects are accessed in short time intervals.

\item
\textit{Recency of accesses}:
    Our traces also have varying recency, i.e., the time interval between consecutive GETs, as shown in Figure-\ref{fig:5_trace_new_analysis}d. 
    Two traces (T15 and T78) have inter-arrival times within a day. In contrast, T65 and T79 show about 10\% of GET intervals falling between one day and one month. In T29, $>$80\% objects are read between one day and up to two months, and the remaining intervals even exceed two months.

\item
\textit{Ratio of GET and PUT operations}:
    Our traces also capture read and write dominant workload patterns.
    Three traces (T65, T78, and T79) are read-heavy, as shown in the Figure~\ref{fig:5_trace_new_analysis}e. The rest (T15 and T29) are write-heavy with 42\% and 30\% of PUTs. Note that T29 has $>$12M requests in total.
\end{itemize}

\noindent
\subsubsection{Multi-region, Multi-cloud workload generation}
\label{eval:multi-cloud-generation}
To address the lack of multi-cloud workloads, we use our five traces to synthetically generate such workloads in three steps.

\noindent
\textbf{Step 1: From one to two regions within a cloud}. We first explore the single-cloud, two-region base and cache setup (as described in Section~\ref{s:problem-twosite}). This setup represents a popular approach~\cite{separate-compute} where data is already located in one region, but the computation is run elsewhere, for instance, due to low resource availability (e.g., geo-distributed model serving service on GPUs). Recall that our traces are from a single region within the IBM cloud. To support this setup, we generate a workload in which PUT operations are directed to the base region and GET operations to the cache region. 

\noindent
\textbf{Step 2: To multiple regions and clouds}.
Next, we generate multi-region and multi-cloud workloads.
Our synthesis of multi-cloud workloads is informed by our conversations with industry experts and their observations, which highlight the following patterns:
\begin{enumerate}[noitemsep=*,leftmargin=*]
    \item \textit{Uniform Workloads (Type A)}:
    Applications like
        networks of IoT sensors~\cite{razzaque2024analysisstatusupdatewireless} and 
        e-commerce websites~\cite{Hristoski_2016}
        have uniformly random access patterns.
    For this workload, we distribute PUTs and GETs 
        randomly across regions and clouds.

    \item \textit{Region-Aware Workloads (Type B)}:
    Applications like 
        satellite image analysis~\cite{googleLandsatData},
        disaster recovery~\cite{disaster-recover}, and 
        cloudburst~\cite{cloud-burst},
        ingest data in one but consume the data from another region. 
    For this workload, we assign unique PUT and GET regions for each object and distribute requests accordingly. 

    \item \textit{Aggregation Workloads (Type C)}:
    Applications that collect data (like regional sales information,
        logs, etc.) at different regions~\cite{lakehouse} but 
        access or analyze this data from a central region.  
    For this workload, we distribute PUTs across regions, allowing data ingestion across regions, and dedicate GETs to a single region.
        
    \item \textit{Replication Workloads (Type D)}: 
    Applications like CDN~\cite{peroni2024endtoendpipelineperspectivevideo}, container registry~\cite{278314}, and
        geo-distributed model serving~\cite{alkassab2023deeprefdeepreinforcementlearning} typically 
        write to a single region and read from
        multiple other regions.
    For this workload, we assign a dedicated  
        PUT region for each object 
        and distribute GETs across other regions.

\end{enumerate}

\noindent
\textbf{Step 3: Multi-cloud workloads}. 
We combine our multi-cloud workloads into
    a single workload (Type E) for a single trace (T65) for our end-to-end experiments with real cloud deployments.
This is necessary as each workload above stores and accesses 6.7 TB of data on average, and it
would cost about 0.2M dollars to evaluate \sys against all workloads and setups in the cloud. 

\shu{R1(W1): In cross-region access scenarios, replicating data across regions may be a common practice in clouds. However, the paper does not leverage the replicas that the cloud inherently provides.}

\begin{figure*}[tbp]
    \centering
    \includegraphics[width=\textwidth]{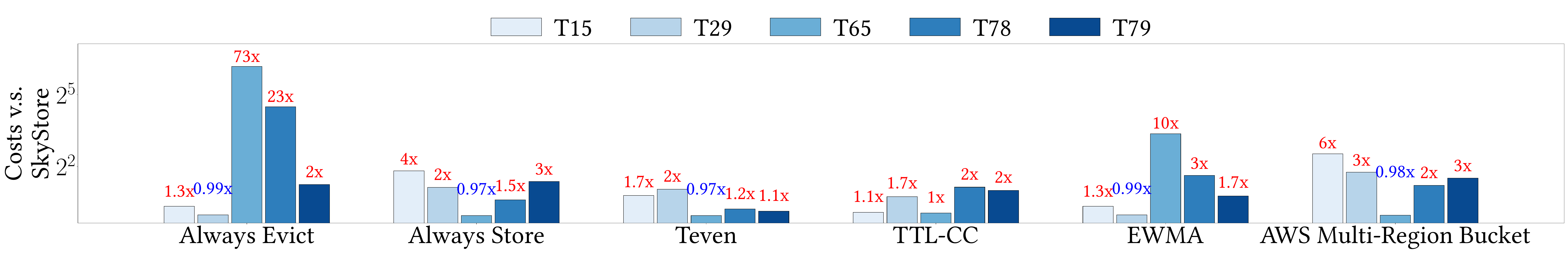}
    \vspace{-2em}

    \caption{\textbf{2-Region, Fixed Base (FB)}: Ratio between baseline cost vs. \sys. On average across traces, \sys is \textbf{1.4-20.0$\times$} cheaper than other baselines.} 
    \label{fig:2sites-update}

\end{figure*}

\begin{table*}
\setlength\extrarowheight{5pt} 
\centering
\resizebox{\textwidth}{!}{%
\Huge
\begin{tabular}{c|cccc|c|cccccc|cc|c|c|ccc}
\hline
 & \multicolumn{5}{c|}{\textbf{Size (\%)}} & \multicolumn{6}{c|}{\textbf{Read Frequency (\%)}} & \multicolumn{3}{c|}{\textbf{Request Arrival}} & \textbf{Recency} & \multicolumn{3}{c}{\textbf{Number of Requests}} \\ \cline{2-19} 
\multirow{-2}{*}{\textbf{\begin{tabular}[c]{@{}c@{}}IBM Trace\\Number\end{tabular}}} & \textbf{\begin{tabular}[c]{@{}c@{}}\# Tiny\\ (\textless{}1kB)\end{tabular}} & \textbf{\begin{tabular}[c]{@{}c@{}}\# Small\\ (1KB-1MB)\end{tabular}} & \textbf{\begin{tabular}[c]{@{}c@{}}\# Medium\\ (1MB-1GB)\end{tabular}} & \multicolumn{1}{c|}{\textbf{\begin{tabular}[c|]{@{}c@{}}\# Large\\ (\textgreater{}1GB)\end{tabular}}} & \textbf{\begin{tabular}[c]{@{}c@{}}Avg.\\ (KB)\end{tabular}} & \textbf{\begin{tabular}[c]{@{}c@{}}One-Hit\\ Wonders\end{tabular}} & \textbf{\begin{tabular}[c]{@{}c@{}}Cold\\ (1--10)\end{tabular}} & \multicolumn{1}{c}{\textbf{\begin{tabular}[c]{@{}c@{}}Warm\\ (10--100)\end{tabular}}} & \textbf{\begin{tabular}[c]{@{}c@{}}Hot\\ (100-1K)\end{tabular}} & \multicolumn{1}{c|}{\textbf{\begin{tabular}[c]{@{}c@{}}Super Hot \\ (\textgreater{}1K)\end{tabular}}} & \textbf{\begin{tabular}[c]{@{}c@{}}Avg.\\ \#GETs\end{tabular}} & \multicolumn{1}{c}{\textbf{\begin{tabular}[c]{@{}c@{}}\%in first 3\\ months\end{tabular}}} & \multicolumn{1}{c|}{\textbf{\begin{tabular}[c]{@{}c@{}}\%in last 4\\ months\end{tabular}}} & \textbf{\begin{tabular}[|c]{@{}c@{}}Avg. GET Tail\\ months\end{tabular}} & \textbf{\begin{tabular}[c|]{@{}c@{}}Avg.\\ days\end{tabular}} & \multicolumn{1}{c}{\textbf{\begin{tabular}[c]{@{}c@{}}GET\\ (\%)\end{tabular}}} & \multicolumn{1}{c}{\textbf{\begin{tabular}[c]{@{}c@{}}PUT\\ (\%)\end{tabular}}} & \multicolumn{1}{c}{\textbf{\begin{tabular}[c]{@{}c@{}}Total\\ (M)\end{tabular}}} \\ \hline
\rowcolor[HTML]{E3EEF9} 
T15 & {0} & {80} & 20 & {0} & 628 & \cellcolor[HTML]{DCEAF7}{\ul \textbf{48}} & 52 & {0} & {0} & \multicolumn{1}{c|}{\cellcolor[HTML]{E3EEF9}{0}} & 3 & 42 & 58 & 2.3 & 0.6 & {57} & {\ul \textbf{43}} & 1.6 \\
\rowcolor[HTML]{B7D4EA} 
T29 & {\ul \textbf{44}} & {56} & {0} & {0} & 3 & 2 & {\ul \textbf{98}} & {0} & {0} & \multicolumn{1}{c|}{\cellcolor[HTML]{B7D4EA}{0}} & 3 & 57 & 43 & 3.5 & {\ul \textbf{41.6}} & 70 & 30 & {\ul \textbf{13}} \\
\rowcolor[HTML]{6AAED6} 
T65 & {31} & {34} & {34} & {0.03} & 1,536 & 2 & 9 & 22 & {\ul \textbf{67}} & \multicolumn{1}{c|}{\cellcolor[HTML]{6AAED6} {\ul \textbf{0.1}}} & {\ul \textbf{93}} & 52 & 48 & 3 & 1.3 & {\ul \textbf{99}} & 1 & 0.3 \\
\rowcolor[HTML]{2E7EBC} 
{\color[HTML]{FFFFFF} T78} & {\color[HTML]{FFFFFF} 0} & {\color[HTML]{FFFFFF} {\ul \textbf{98}}} & {\color[HTML]{FFFFFF} 2} & {\color[HTML]{FFFFFF} 0} & {\color[HTML]{FFFFFF} 578} & {\color[HTML]{FFFFFF} 6} & {\color[HTML]{FFFFFF} 31} & {\color[HTML]{FFFFFF} {\ul \textbf{51}}} & {\color[HTML]{FFFFFF} 11} & \multicolumn{1}{c|}{\cellcolor[HTML]{2E7EBC}{\color[HTML]{FFFFFF} \ul \textbf{0.1}}} & {\color[HTML]{FFFFFF} 26} & {\color[HTML]{FFFFFF} 22} & {\color[HTML]{FFFFFF} \ul \textbf{78}} & {\color[HTML]{FFFFFF} 0.8} & {\color[HTML]{FFFFFF} 2.6} & {\color[HTML]{FFFFFF} 95} & {\color[HTML]{FFFFFF} 5} & {\color[HTML]{FFFFFF} 2.4} \\
\rowcolor[HTML]{084A91} 
{\color[HTML]{FFFFFF} T79} & {\color[HTML]{FFFFFF} 0} & {\color[HTML]{FFFFFF} 40} & {\color[HTML]{FFFFFF} {\ul \textbf{60}}} & {\color[HTML]{FFFFFF} {\ul \textbf{0.35}}} & {\color[HTML]{FFFFFF} \ul \textbf{48,386}} & {\color[HTML]{FFFFFF} 17} & {\color[HTML]{FFFFFF} {61}} & {\color[HTML]{FFFFFF} 22} & {\color[HTML]{FFFFFF} 0} & \multicolumn{1}{c|}{\cellcolor[HTML]{084A91} \color[HTML]{FFFFFF}{0}} & {\color[HTML]{FFFFFF} 9} & {\color[HTML]{FFFFFF} {\ul \textbf{60}}} & {\color[HTML]{FFFFFF} 40} & {\color[HTML]{FFFFFF} {\ul \textbf{4.1}}} & {\color[HTML]{FFFFFF} 8.3} & {\color[HTML]{FFFFFF} 89} & {\color[HTML]{FFFFFF} 11} & {\color[HTML]{FFFFFF} 0.1}
\end{tabular}%
}
\vspace{2pt}
\caption{IBM Trace Characteristics: each trace with characteristics (size, read frequency, request arrival, recency, number of requests) highlighted with bold and underscore.}
\label{tbl:trace_characteristics}
\vspace{-2em}
\end{table*}

\subsection{Deployment Settings and Baselines}\label{eval:metrics_baselines_deploy}
We evaluate \sys's policy in two deployment settings (Section~\ref{eval:deploy}) and compare against several baselines (Section~\ref{eval:baselines}).

\subsubsection{Deployment settings and metrics}
\label{eval:deploy} 
Policies can operate in one of two modes: fixed base (FB) and free placement (FP). \sys assumes FB mode by default, where each object has a fixed, non-evictable base region. In FP mode, any replicas can be evicted, but at least one always remains. Since our closest related work, SPANStore operates only in FP mode, \sys supports both modes and compares against SPANStore in FP mode. We assume read-after-write consistency with version enabled, where each read accesses the latest data version.
\newline 
\noindent
\textbf{Multi-cloud deployment settings}. Our multi-cloud deployments span across AWS S3, Azure Blob Storage, and GCS clouds. We run 3-region\footnote{\path{aws:us-east-1}, \path{azure:eastus}, \path{gcp:us-east1-b}}, 6-region\footnote{\path{aws:us-east-1}, \path{aws:us-west-2}, \path{azure:eastus}, \path{azure:westus}, \path{gcp:us-east1-b}, \path{gcp:us-west1-a}}, and 9-region\footnote{\path{aws:us-east-1}, \path{aws:us-west-2}, \path{aws:eu-west-1}, \path{azure:eastus}, \path{azure:westus}, \path{azure:wasteurope}, \path{gcp:us-east1-b}, \path{gcp:us-west1-a}, \path{gcp:europe-west1-b}} experiments where we select 1, 2, and 3 regions from each cloud provider, respectively. 
\newline 
\noindent
\textbf{Metrics}.
We compare \sys against other baselines on cost and latency metrics. We measure the total monetary cost of running a workload based on the standard storage offerings and bi-directional network costs between cloud regions.
We also measure the average, p90, and p99 latency for GET and PUT requests. 

\subsubsection{Baselines}
\label{eval:baselines}
We compare against the following baselines:
\begin{itemize}[noitemsep=*,leftmargin=*]
    \item 
    \textbf{Always Store / Always Evict} policy always replicates objects to regions where GET is initiated and never evicts, or stores each object in a single storage location and never replicates.
    
    \item
    \textbf{TTL-based Eviction} policies include 
    (a) $\tev=\teven$ (Section~\ref{eqn:teven}),
    (b) TTL-CC~\cite{TTLbasedCC2019}, a dynamic policy that stochastically sets TTL based on the cached object's behavior. \shu{Consider a variant of TTL-CC such that it sets TTL for each object, based on its access pattern, rather than setting the "TTL of all objects". The expectation is that the proposed mechanism, GlobalState, will outperform per-object TTL-CC in some workloads.
}

    \item
    \textbf{Clairvoyant Greedy Policy (CGP)} (Section~\ref{sec:clairvoyant}) is an oracle that decides to store or evict given future access times of each object. CGP is cost-optimal in the two-region setup. 

    \item
    \textbf{EWMA}
    uses an Exponentially Weighted Moving Average \cite{ewma} to predict the next access time per object and chooses whether to evict it accordingly. We set the decay factor $\alpha$ to be 0.5. 

    \item
    \textbf{SPANStore} is a multi-cloud replication policy~\cite{spanstore} that replicates objects each hour to minimize access costs. SPANStore does not fix a storage location and hence, we evaluate it only in the free placement (FP) mode. 

    \item
    \textbf{Industrial Baselines} include AWS Multi-Region Buckets~\cite{aws-access-point} (and similarly, GCP Multi-Region Bucket~\cite{gcp-multi-region-bucket}) and JuiceFS~\cite{juicefs}. Upon PUT, an object is asynchronously replicated to the pre-configured secondary region(s). We evaluate AWS in a two-region setup and JuiceFS in a multi-cloud setup, assuming the object is replicated to all other regions. \shu{R2 / R1: Expanding the comparison to include a wider range of industry-standard cost-optimization solutions would strengthen the paper. I wonder whether including benchmarks like Google Cloud Storage, Microsoft Azure Blob Storage, and solutions optimized for multi-cloud setups could offer a more comprehensive view of where GlobalStore stands in the current landscape.
}
\end{itemize}


\subsection{Single-Cloud Two Regions: Base and Cache}
\label{eval:two-region}

We now evaluate \sys in a two-region base and cache setup and showcase its merit as a standalone caching policy.
\sys consistently maintains low costs across traces, while its alternatives have low or comparable costs in specific cases and become prohibitively more expensive in others. 
On average, \sys has \textbf{1.4--20$\times$ lower costs} compared to six baselines (Section-\ref{eval:baselines}) across five traces, as shown in Figure~\ref{fig:2sites-update}.

\textit{AlwaysEvict} is effective on one-hit-dominant traces, as it avoids unnecessary storage costs for objects never accessed again. For instance, in T15 where 48\% of objects are accessed only once (Table~\ref{tbl:trace_characteristics}), AlwaysEvict incus only 30\% higher costs than \sys. 
It results in slightly higher network costs for the remaining cold objects accessed more than once in this trace.
In traces like T29, where there is longer average recency between GETs (beyond $\teven$), the cost of storing data outweighs the cost of fetching it again on the subsequent access. 
In such cases, AlwaysEvict can even outperform \sys by 1\% , as \sys reactively caches objects and requires time to adjust to a lower TTL.
On the other hand, on traces like T65 and T78 with warm and hot accesses and shorter access recency, AlwaysEvict incurs 23--73$\times$ higher costs due to repeated network transfers on reads. 
Surprisingly, in T79 where 89\% of objects are one-hits or cold,
    AlwaysEvict still costs 2$\times$ more than \sys. This is primarily due to the large average object size (48MB), which amplifies the network cost penalties from cache misses.
On average, 
    \sys is 20$\times$ more cost-effective than AlwaysEvict.

\textit{Always Store} replicates objects on GETs and exhibits behavior that almost contrasts with AlwaysEvict.
On traces with lots of hot objects such as T65, AlwaysStore outperforms \sys by 3\%, as \sys may evict a few hot objects and incur higher network costs during the initial histogram warmup phase.
However, on traces with more infrequent or sporadic access patterns like T29 and T15,  AlwaysStore incurs 2--4$\times$ higher costs than \sys. 
Interestingly, on traces with frequent repeated reads, such as T79 where 80\% of the objects are accessed multiple times, Always Store remains 3$\times$ more expensive than \sys. This is because 60\% of the objects in T79 have a GET tail longer than 4.1 months ( as seen in Table~\ref{tbl:trace_characteristics}), which causes AlwaysStore to retain objects long after their last access. \sys evicts unaccessed objects earlier and outperforms AlwaysStore by 1.5--3$\times$ on T78 and T79. On average, \sys is 2.2$\times$ cheaper than Always Store.




\textit{$\teven$} is a static TTL-based policy (Section~\ref{s:teven-polcy}) that stores objects until re-fetching them becomes less expensive and balances storage with network costs. 
    In our setup, the TTL for $\teven$ policy is one month, calculated as the ratio between average network cost and standard S3 prices across 22 AWS regions.
$\teven$ performs well when all GETs occur within a month (as in T65), enabling timely evictions and slightly outperforming \sys in this case.
However, for
    infrequent accesses like in T15 and T29, it stores objects for a full one-month TTL, leading to 1.7-2$\times$ higher costs compared to \sys.
In traces with moderate access frequency and short recency (like T15 and T29), $\teven$ strikes a reasonable balance. However, \sys still outperforms it by 1.4$\times$ as, unlike $\teven$, \sys is aware of object access patterns and can reduce network costs with its adaptive TTL.
On average, \sys is 1.4$\times$ more cost-effective than $\teven$.

    
    

\textit{TTL-CC} policy\cite{TTLbasedCC2019} computes TTLs stochastically based on cache hits, assuming a Poisson distribution, and dynamically updates the TTL of all objects. 
This policy's cost is within 10\% of the total cost of \sys for traces with hot- or one-hit-dominant objects (like T15 and T65).
However, for mixed traces with warm and cold objects like T78 and T79, TTL-CC has 2$\times$ higher cost than \sys.
TTL-CC also tends to store sporadically accessed objects for longer in T29 and incurs 1.7$\times$ higher cost.
In summary, TTL-CC results emphasize that dynamic TTL-based policies are a better fit for cloud applications. However, access patterns in the cloud are more complex than Poisson distributions. Overall, \sys is more cost-efficient than TTL-CC by 1.6$\times$ on average.



\textit{EWMA} predicts object access times using exponentially weighted moving averages and stores objects with shorter access times. 
This policy can quickly evict one-hits and cold objects and reduces storage costs by 0.99--1.3$\times$ for traces T29 and T15, respectively. 
However, it carries this aggressive eviction strategy over to traces with hot and super-hot objects (T65, T78, and T79) and incurs 1.7--10$\times$ higher costs. Fine-tuning EWMA policy parameters, such as the decay factor, could potentially reduce these overheads.
On average, EWMA is 3.5$\times$ more expensive than \sys.


\textit{AWS Multi-Region Bucket}~\cite{aws-access-point} and similar commercially-available services behave like AlwaysStore but proactively replicate data on writes rather than reads. This leads to higher storage costs when GET appears later; in traces T15, T29, and T78, on average, objects are accessed 1--1.5 months after they are written, incurring 2--6$\times$ higher cost than \sys. For traces with more immediate reads (like T65 and T79), AWS multi-region buckets incur 0.98--3$\times$ higher costs than \sys.
Overall, AWS Multi-Region Bucket is 3.1$\times$ more expensive than \sys on average.



We also compare \sys to CGP, an oracle with optimal cost policy (Table~\ref{tab:cost_vs_optimal}). On average, \sys operates within 15\% of optimal, while others incur 1.6--22$\times$ higher costs. \sys incurs higher costs as it aggregates statistics at bucket granularity. \sys performs 1.2--1.3$\times$ worse than CGP for traces (T78, T79) with a mixed distribution of access frequencies and recency where aggregate statistics become less accurate. Note that $\teven$ is empirically within 2$\times$ of optimal cost, as proven in Section~\ref{s:teven-polcy}.


\begin{table}[ht]
\centering
\small
\setlength{\tabcolsep}{5pt} 
\renewcommand{\arraystretch}{1} 
\begin{tabular}{lcccccc}
\hline
\textbf{Policy}       & \multicolumn{6}{c}{\textbf{Cost vs. Optimal}} \\
                      & \textbf{T15} & \textbf{T29} & \textbf{T65} & \textbf{T78} & \textbf{T79} & \textbf{Avg} \\
\hline
Always Evict          & 1.4 & 1.0 & 77.5 & 27.8 & 3.1 & 22.15 \\
Always Store          & 3.9  & 2.3  & 1.0  & 1.9  & 3.4  & 2.49  \\
Teven                 & 1.9  & 2.2  & 1.0  & 1.4  & 1.5 & 1.59  \\
TTL-CC                & 1.2  & 1.7  & 1.1  & 2.7  & 2.7 & 1.87  \\
TTL-CC-obj                & 1.5  & 1.0  & 7.5  & 7.2  & 2.2 & 3.88  \\
EWMA                  & 1.4  & 1.0  & 11.0  & 3.8  & 2.3  & 3.90  \\
AWS Multi-Region Bucket & 6.3  & 3.5  & 1.0  & 2.8  & 3.8  & 3.49  \\

\cellcolor{Gray}\sys   & \cellcolor{Gray}1.1  & \cellcolor{Gray}1.0  & \cellcolor{Gray}1.1  & \cellcolor{Gray}1.2  & \cellcolor{Gray}1.3  & \cellcolor{Gray}1.14  \\ 

\hline
\end{tabular}
\vspace{0.3em}
\caption{Two-Region Base and Cache: Cost vs. Optimal across individual traces and their average. \sys is, on average, within 14\% of optimal cost.}
\label{tab:cost_vs_optimal}
\vspace{-3em}
\end{table}

\begin{table}
\centering
\small
\setlength{\tabcolsep}{1.8pt} 
\renewcommand{\arraystretch}{0.9} 
\begin{tabular}{lccccc} 
\toprule
\textbf{Policy} & \makecell{\textbf{Type A} \\ (Uniform)} & \makecell{\textbf{Type B} \\ (Region)} & \makecell{\textbf{Type C} \\ (Aggregation)} & \makecell{\textbf{Type D} \\ (Replication)} & \textbf{Average} \\
\midrule
Always Evict & 9.3 & 29.8 & 24.0 & 10.4 & 18.4$\times$ \\
Always Store & 1.8 & 1.7 & 1.7 & 1.9 & 1.8$\times$ \\
$\teven$ & 1.3 & 1.3 & 1.3 & 1.3 & 1.3$\times$ \\
TTL-CC & 1.7 & 1.2 & 1.3 & 1.8 & 1.5$\times$ \\
EWMA & 2.9 & 4.9 & 4.4 & 3.0 & 3.8$\times$ \\
JuiceFS & 4.8 & 1.9 & 1.9 & 4.8 & 5.7$\times$ \\
\bottomrule
\end{tabular}
\vspace{0.8em}
\caption{3-Region Fixed Base: baseline cost over \sys ($\times$), averaged across traces and across workload types. On average, \sys is 1.3 to 18.4$\times$ cheaper than six other baselines.}
\label{tab:cost_comparison_3_region_avg}
\vspace{-2em}
\end{table}

\subsection{Multi-Cloud: 3 Regions across 3 Clouds}
\label{eval:three-region}

We extend our evaluation to a multi-cloud setup with three regions across three clouds (Section~\ref{eval:deploy}). We use four workloads, i.e., uniform, region-aware, aggregation, and replication workloads (Section~\ref{eval:multi-cloud-generation}).
Across these workloads, \sys consistently achieves lower costs compared to other baselines by \textbf{1.3--18.4$\times$} on average. Table~\ref{tab:cost_comparison_3_region_avg} summarizes the baseline's cost over \sys's, and averages it across traces and workload types. 
We compare \sys with JuiceFS instead of AWS Multi-Region Bucket as the latter does not operate across clouds. All policies in this experiment are run in the fixed base (FB) mode.

At a high level, 3-region multi-cloud results largely mirror the 2-region setup. Major trends across traces remain the same, but the absolute cost improvements differ from the two-region setup across each workload. This is primarily due to higher (1.8$\times$) network fees in multi-cloud compared to a single-cloud setup. We highlight and explain outlier trends in costs for \sys and other baselines.

\textit{AlwaysEvict} is 9.3--29.8$\times$ more expensive than \sys on average. 
AlwaysEvict has higher costs (29.8$\times$ and 24.0$\times$) in region-aware and aggregation workloads; for read-heavy traces like T65, it can be 120$\times$ worse due to high cross-cloud network fees.
AlwaysEvict performs slightly better (9.3$\times$ and 10.4$\times$) than \sys on uniform and replication workloads. 
    As GETs per object are distributed, policies have uniformly high network costs of cold misses in each region, narrowing their performance gap.

\textit{EWMA} has similar costs to AlwaysEvict except that it retains objects for slightly longer in read-heavy traces, costing 2.9 -- 4.9$\times$ more than \sys on average. 

\begin{figure*}[tbp]
    \centering
    \includegraphics[width=.95\textwidth]{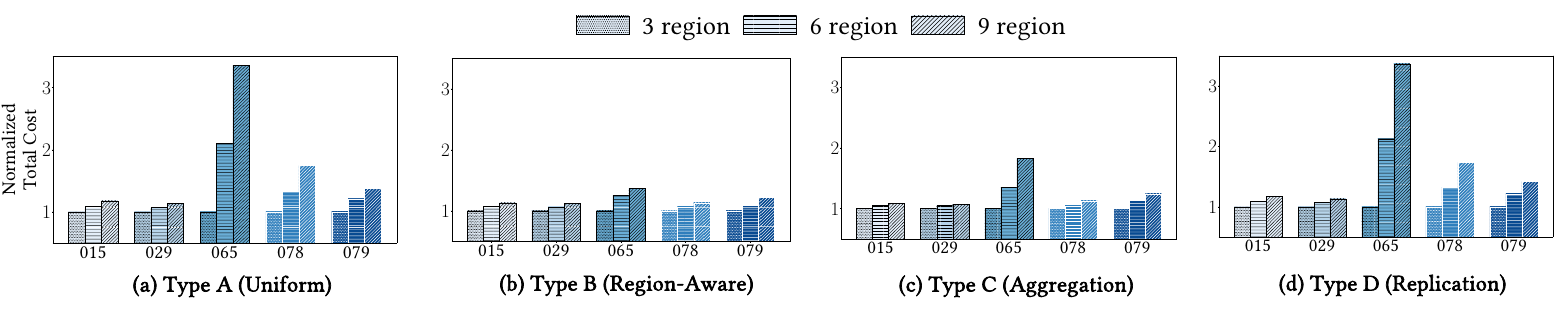}
    \vspace{-1em}
    \caption{\textbf{\sys Total Cost Normalized over 3-Region on 3, 6, and 9 Regions across Workloads A-D: \sys costs remain similar when scaling to more regions.}}
    \label{fig:3-6-9-scalability}
    \vspace{-1em}
\end{figure*}

\textit{AlwaysStore} and \textit{JuiceFS} are 1.7--1.9$\times$ and 1.9--4.8$\times$ more expensive than \sys on average, respectively.  
Decreasing access frequency and increasing recency reduces caching benefits. 
Thus, AlwaysStore incurs low costs (1.7$\times$ of \sys) in region-aware and aggregation workloads and higher costs (1.8 -- 1.9$\times$) with uniform and replication workloads.
Surprisingly, AlwaysStore beats \sys by 4-6\% on read-heavy traces (T65) as \sys incurs higher network costs during initial metadata warmup periods.
In contrast, JuiceFS has 2.7$\times$ higher costs than AlwaysStore on average because JuiceFS proactively replicates objects to all regions on PUTs and incurs high costs for infrequently read objects.
However, if read locations are predictable, such as region-aware and aggregation workloads, JuiceFS is auto-configured to replicate to specific regions and incurs similar costs as AlwaysStore.

\textit{TTL-CC} costs 1.2--1.8$\times$ more than \sys on average and adjusts objects TTLs based on their cache hit rate. In general, TTL-CC has lower costs (1.2--1.3$\times$ vs. \sys) for region-aware and aggregation workloads, but higher costs (1.7--1.8$\times$) for uniform or replication workloads. 
As an exception, TTL-CC incurs 1.6$\times$ lower costs (for trace T79) in uniform and replication workloads than in region-aware and aggregation workloads. Since cold objects (61\% of objects in T79),
    have even fewer accesses when GETs are distributed across regions,
    TTL-CC learns this pattern quickly and sets shorter TTLs.
However, it adapts slowly when reads are concentrated in specific locations,
    evicts relatively warmer objects and incurs higher network costs.

The $\teven$ policy incurs low costs consistently, i.e., 1.3$\times$ higher than \sys on average. It balances storage and network costs in multi-cloud setups and maintains low costs across different request distributions. $\teven$ has up to 1.7$\times$ higher costs than \sys as it stores objects even if they are not read in the future.

\subsection{Scalability: To 9 Regions across 3 Clouds}
\label{eval:nine-region}

\begin{table}[t]
\centering
\small
\begin{tabular}{lcccccc}
\toprule
\textbf{Policy} & \multicolumn{2}{c}{\textbf{3-Region (FB)}} & \multicolumn{2}{c}{\textbf{6-Region (FB)}} & \multicolumn{2}{c}{\textbf{9-Region (FB)}} \\
\cmidrule{2-7}
& \textbf{Avg} & \textbf{Std Dev} & \textbf{Avg} & \textbf{Std Dev} & \textbf{Avg} & \textbf{Std Dev} \\
\midrule
\textit{Always Evict} & 18.4 & 31.6 & 15.0 & 27.0 & 12.9 & 23.9 \\
\textit{Always Store} & 1.8 & 0.7 & 2.0 & 0.8 & 2.1 & 0.8 \\
\textit{Teven} & 1.3 & 0.3 & 1.4 & 0.3 & 1.4 & 0.3 \\
\textit{TTL-CC} & 1.5 & 1.0 & 1.3 & 0.5 & 1.4 & 0.7 \\
\textit{EWMA} & 3.8 & 4.7 & 3.2 & 3.4 & 3.1 & 3.4 \\
\textit{JuiceFS} & 3.4 & 2.9 & 7.3 & 7.2 & 8.6 & 10.9 \\
\midrule
\textbf{Policy} & \multicolumn{2}{c}{\textbf{3-Region (FP)}} & \multicolumn{2}{c}{\textbf{6-Region (FP)}} & \multicolumn{2}{c}{\textbf{9-Region (FP)}} \\
\cmidrule{2-7}
& \textbf{Avg} & \textbf{Std Dev} & \textbf{Avg} & \textbf{Std Dev} & \textbf{Avg} & \textbf{Std Dev} \\
\midrule
\textit{Always Evict} & 11.9 & 21.1 & 11.4 & 22.4 & 11.4 & 23.4 \\
\textit{Always Store} & 1.3 & 0.2 & 1.4 & 0.2 & 1.5 & 0.2 \\
\textit{JuiceFS} & 1.7 & 0.6 & 3.0 & 1.8 & 3.5 & 2.8 \\
\textit{SPANStore} & 1.4 & 0.2 & 1.5 & 0.2 & 1.6 & 0.3 \\
\bottomrule \\
\end{tabular}%
\caption{3, 6, 9-Region, 5 traces, Type A-D, Fixed Base (FB) and Free Placement (FP): Average and standard deviation of cost of baselines over \sys. Scaling to 9 regions, \sys can still achieve 1.4 to 12.9$\times$ and 1.5 to 11.4$\times$ cheaper cost on average than other baselines in FB and FP modes, respectively.}
\vspace{-2em}
\label{table:avg_std_combined}
\end{table}

We now evaluate how \sys's cost savings scale in the multi-cloud setup with an increasing number of regions.
Across three clouds,
    we compare cost savings of \sys in three, six, and nine regions relative to AlwaysEvict, AlwaysStore, and commercial/academic baselines like JuiceFS and SPANStore.
Note that we evaluate SPANStore only in FP mode as it does not support FB mode. Across 9 regions, \sys is \textbf{1.4--12.9$\times$} and\textbf{ 1.5--11.4$\times$} more cost-efficient than other baselines in FB and FP modes, respectively.



Table~\ref{table:avg_std_combined} summarizes how scaling affects baseline costs relative to \sys, so lower cost relative to \sys showcases better scalability.
\sys remains consistent and incurs low costs when scaling regions.
AlwaysEvict and EWMA (in FB mode) incur lower costs on increasing the number of regions from 3 to 9 (18.4 to 12.9$\times$, 3.8 to 3.1$\times$, respectively).
On the other hand, AlwaysStore and JuiceFS incur higher costs (1.8 to 2.1$\times$, 3.4 to 8.6$\times$, respectively) compared to \sys as regions increase. This is primarily because the number of data replicas is proportional to the number of regions, and these policies incur high storage costs from extensive replication. Recall that JuiceFS proactively replicates data to all regions on PUT requests and pays for higher storage and network costs as regions scale.  
Both $\teven$ and TTL-CC remain fairly consistent (1.3-1.4$\times$ and 1.3-1.5$\times$), and show slight fluctuations in relative cost compared to \sys when scaling from 3 to 9 regions.

\sys and other policies have relatively lower costs in FP relative to FB mode as they incur no additional costs for the base region's storage. SPANStore has comparable costs
to AlwaysStore as it does not effectively evict objects that remain unread for long time intervals.
SPANStore incurs even higher costs for traces with a majority of one-hits and cold objects (like traces T29 and T79). In our evaluation, SPANStore's solver has access to an oracle with knowledge of workloads and showcases its costs in the best case. Across workloads and traces, \sys is 1.4--1.6$\times$ more cost efficient than SPANStore on average, with 9 regions across 3 clouds.

\sys's cost savings scale from 3 to 9 regions across workload types and traces in FB mode, as seen in Figure-\ref{fig:3-6-9-scalability}.
On region-aware and aggregation-workloads (Figures~\ref{fig:3-6-9-scalability}b,~\ref{fig:3-6-9-scalability}c), \sys has minimal cost variations with more regions. In these workloads, GETs of objects are concentrated in a single region independent of the number of regions. 
As an exception, aggregation workloads for trace T65 experience higher costs on scaling to 6 and 9 regions due to a particular cloud region (\texttt{aws:us-east-1}), which has higher network ingress costs from all other regions. 
For uniform and replication workloads (Figure~\ref{fig:3-6-9-scalability}a, ~\ref{fig:3-6-9-scalability}d), \sys's cost remains relatively stable as regions increase. This trend is evident in traces with cold objects (T15, T29, and T79), where scaling to 9 regions yields similar costs. However, for traces with warm and hot objects \sys's cost increases with the number of regions (like 1.5$\times$ and 1.2$\times$ for traces T65 and T78) as GETs are distributed across more regions which makes previously warm objects now colder, and increases network costs from evicting such objects.

\subsection{Multi-Cloud: End-to-End Benchmark}
\label{eval:end-to-end}


\begin{table}[ht]
\centering
\footnotesize
\setlength{\tabcolsep}{4pt} 
\renewcommand{\arraystretch}{1.2} 
\begin{tabularx}{\linewidth}{l*{3}{>{\centering\arraybackslash}X}*{3}{>{\centering\arraybackslash}X}p{1.2cm}p{1.1cm}} 
\toprule
\textbf{Policy} & \multicolumn{3}{c}{\makecell{\textbf{GET} \\ \textbf{Latency (ms)}}} & \multicolumn{3}{c}{\makecell{\textbf{PUT} \\ \textbf{Latency (ms)}}} & {\makecell{\textbf{GET Lat.} \\ \textbf{vs. AS}}} & {\makecell{\textbf{Cost (\$)} \\ \textbf{vs. AS}}} \\
\cmidrule(lr){2-4} \cmidrule(lr){5-7} 
 & \textbf{Avg} & \textbf{P90} & \textbf{P99} & \textbf{Avg} & \textbf{P90} & \textbf{P99} &  & \\
\midrule
Always Store & 172 & 235 & 340 & 840 & 562 & 784 & 1.00$\times$ & 1.00$\times$ \\
Always Evict & 278 & 440 & 762 & 800 & 507 & 715 & 1.61$\times$ & 76.78$\times$ \\
\sys  & 184 & 230 & 408 & 822 &  520 & 782 & 1.06$\times$ & 1.05$\times$ \\
\bottomrule
\end{tabularx}
\vspace{0.5em}
\caption{End-to-End System Evaluation on T65. \sys has comparable latency and incurs low costs on real multi-cloud deployments with 3 regions across 3 clouds.}
\label{tab:end_to_end_result}
\vspace{-2em}
\end{table}

We now discuss the end-to-end cost and latency of \sys against
      AlwaysStore and AlwaysEvict baselines for a multi-cloud workload (Type E) on a single trace (T65) due to the prohibitively high cost (2M dollars) of evaluating all workloads and configurations (Section-\ref{eval:multi-cloud-generation}).
We run \sys and baselines on 3 regions across 3 clouds.
As seen in Table~\ref{tab:end_to_end_result},
     \sys has comparable average and p99 latency as AlwaysStore, with 3\% higher average GET latency due to its metadata overheads from maintaining per-bucket statistics as histograms and 
     periodically updating them in the background.
Note that PUT latency (average and p99) is similar across policies, as writes are handled locally.
However, AlwaysEvict avoids caching objects and observes 1.6$\times$ higher average GET latency than \sys and AlwaysStore.

\sys and AlwaysStore policies incur low costs in real cloud deployments and align with our cost simulations from 3-region 3-cloud experiments (Section~\ref{eval:three-region}).
AlwaysEvict baseline policy incurs up to \textbf{75$\times$ higher costs}
and has higher end-to-end latency in comparison to \sys and AlwaysStore.

\subsection{Discussions: Overheads \& Trade-offs}
\label{eval:overheads}

\subsubsection{Cost overheads}
The monetary cost of operating \sys comes from two components: the S3-proxy and the metadata server. The S3-proxy is run as a client-side library to incur no additional costs. The metadata server, which manages policy decisions and namespace mappings, is hosted on a standalone VM. In our evaluation, we use a m5d.8xlarge instance costing \$1.81 per hour.
This is analogous to the operational costs of cloud service providers.
\vspace{-0.5em}
\subsubsection{System Overheads}
\label{s:system-overhead}
\begin{figure}
    \includegraphics[width=.85\linewidth]{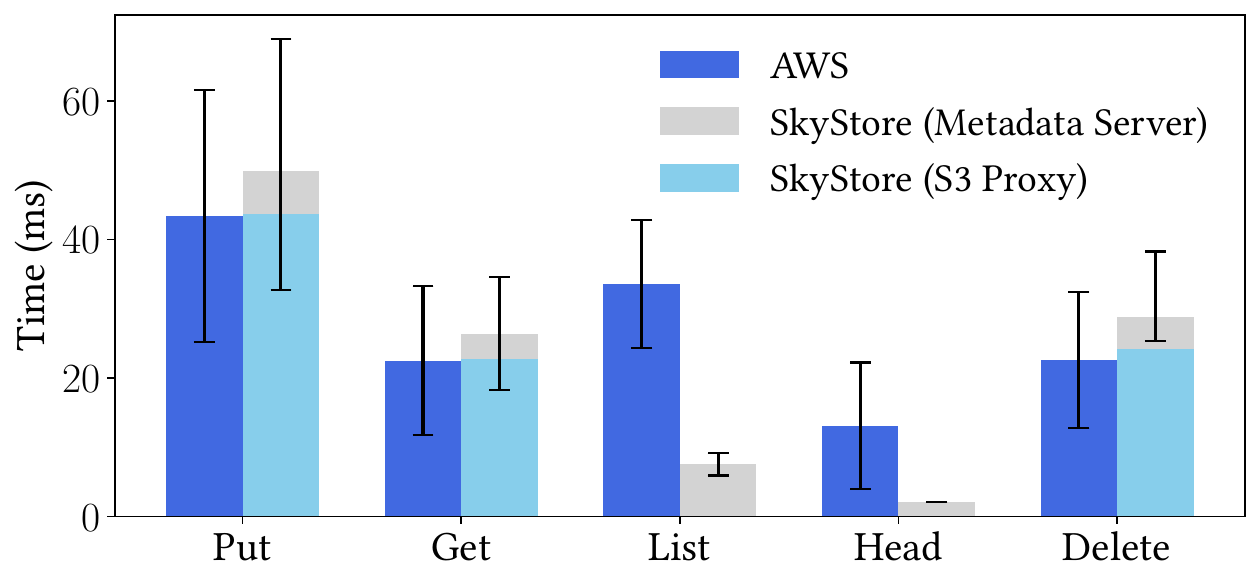}
    \vspace{-1em}
    \caption{\sys System Performance v.s. AWS APIs: client, server, and object store on region \textit{aws:us-east-2}.}
    \label{fig:proxy-overhead}
    \vspace{-1.5em}
\end{figure}

We evaluated \sys's system overhead using the JuiceFS benchmark, which tests operations on 10K objects (128 KB each) across put, get, list, head, and delete actions. As shown in Figure~\ref{fig:proxy-overhead}, \sys adds less than 10\% overhead for put and get operations; this overhead is mainly from additional round-trips to the metadata server. \sys improves the latency of list and head operations by up to 3.4$\times$ with its centralized control place. \sys's S3-proxy is stateless and easily scalable by deploying multiple proxies per client VM. In \sys, scaling the metadata server is straightforward as it does not store actual object data. We host the metadata server's database on a single VM, but a geo-distributed database like Google Spanner~\cite{spanner-cloud} could potentially handle more requests and scale efficiently. Exploring this is left for future work. 
\shu{R2 Stress testing: Adding detailed testing on GlobalStore’s scalability for extremely large data volumes and assessing the latency impact of data transfers under peak conditions would help address concerns about performance trade-offs. This could involve simulations of scaling scenarios that showcase latency effects and help pinpoint possible optimizations.
}
\vspace{-0.5em}
\subsubsection{Overheads with scaling regions and buckets.}
\sys is designed to scale effectively with the increasing number of regions and buckets. Histograms are generated periodically (once or twice a day) for each bucket. The complexity of generating histograms is linear in size. For each bucket, the system calculates point-to-point access patterns. If there are ten regions, this results in $10^2=100$ edges per bucket. For 1000 buckets, this scales to 100,000 edges, which becomes manageable with daily or periodic updates.
\shu{R2: A more nuanced analysis of GlobalStore under different access patterns (e.g., bursty, highly localized, and highly distributed patterns) would improve understanding of its adaptability. Testing GlobalStore under workloads with high variability in access frequency could reveal its robustness and limitations, especially for latency-sensitive applications.}

\vspace{-0.5em}
\subsubsection{How does \sys incorporate latency considerations?}

We illustrate a simple example of latency consideration (Section~\ref{sec:latency}). Consider a scenario where storing an object costs \$0.026/GB per month, with a \$0.02/GB network egress fee for fetching from another region. The default TTL of 0.77 months (from $\teven = N/S$) means we keep the object cached if it’s accessed within 0.77 months. Now, if the user is willing to pay an extra \$0.005/GB for faster access, we check if extending the TTL to 1 month is worthwhile. The additional storage cost for the extra 0.23 months is \$0.006/GB, but the user performance value is only \$0.005/GB, making it not worth extending the TTL to 1 month. However, if the TTL is reduced to 0.5 months, the storage cost saved is greater than the network cost for refetching, and the added performance value could justify shorter caching.
For high-latency tolerance, the user may opt for lower $\uval$, whereas time-sensitive applications may justify a higher $\uval$ for keeping objects cached longer.
We explore this further as part of our future work.
\shu{R2: A use-case study: Providing users with additional configuration options or granular controls within the TTL-based placement policy could make GlobalStore more customizable. This could allow users to specify retention rules based on data types or compliance requirements, enhancing GlobalStore’s applicability for regulated industries.
}

\noindent

\section{Related Work}

\textbf{Geo-distributed Cloud Storage.} Existing commercial offerings are mostly not multi-cloud, and require manual placement decisions. AWS \cite{aws-replication} and GCP multi-region buckets\cite{gcp-multi-region-bucket} are primarily designed for disaster recovery and do not support auto-replication based on workload patterns. Cloudflare's global object store~\cite{cloudflare-r2} does not disclose implementation details and also requires manual configuration. 
%
Volley and Nomad \cite{volley, nomad} do instead optimize data placement for geo-distributed applications. However, they focus primarily on moving data across multiple data centers to minimize access latency and thus ignore the monetary aspect of cloud storage. SPANStore~\cite{spanstore}, a multi-cloud replication system, does consider financial costs but optimizes object placement periodically only. Moreover, it performs proactive replication on writes and does not consider eviction and replication costs. It expects apriori workload knowledge and thus cannot react to evolving workload patterns. \newline 
\textbf{Traditional Caching Algorithms.} A range of traditional cache eviction algorithms have been developed based on object statistics such as recency, frequency or size (e.g., LRU, LFU, GDSF, FIFO). However, these algorithms consider \textit{cache space} as the primary driver for eviction. Object must thus be \textit{ranked}; objects with the lowest ranking are evicted when the cache becomes full. In contrast, the cache in multi-cloud is not constrained by size but by cost. Each caching decision can thus be made independently for each object.  
\newline 
\textbf{TTL-based Caching and Cloud Caching.}
TTL-based cache eviction are popular eviction strategies with average performance similar to LRU~\cite{TTL2003,TTL2012}. TTL-based approaches have been used for cloud caching, 
setting a TTL per cache item according to object read frequency~\cite{tokeep2014}. Tokeep et al. keeps an item for $\teven$ time and evicts it if it has shown no hits. This is equivalent to the $\teven$-policy we evaluate. Carra et al.~\cite{TTLbasedCC2019} offers an approach closer to our two-site approach of a single dynamic TTL for all items in a workload. 
They use a stochastic approach that modifies the TTL by tracking hits of each new item in the cache. Both prior works assume that object reads occur according to set distributions. 
In contrast, \sys policy adapts to changing workload distributions.

\section{Conclusion}
This paper explores the problem of designing a cost-optimized object store across regions and clouds. We propose a TTL-based cost-aware replication policy in the multi-region and multi-cloud setting and build a global object store as an overlay that sits on top of multiple existing cloud services. Our evaluation shows that \sys can achieve up to 6$\times$ cost savings over state-of-the-art baseline policies and systems in a real cloud setup. 
\newpage

\bibliographystyle{plain}

\bibliography{reference}

\end{document}